\pdfoutput=1

\documentclass[preprint,3p]{elsarticle}

\usepackage{lineno,hyperref}
\usepackage{bm}
\usepackage{amsmath}
\usepackage{mathtools}
\usepackage{color}
\usepackage{algorithm}
\usepackage{algorithmic}
\usepackage{multirow}
\modulolinenumbers[5]

\usepackage{amssymb}

\usepackage{subfigure}
\usepackage{array}
\newcolumntype{C}[1]{>{\centering\arraybackslash}p{#1}}

\usepackage{bigints}

\usepackage{bbding}

\usepackage{units}

\usepackage{xcolor}

\newcommand{\Div}[1]{\nabla \cdot {#1}}
\newcommand{\Grad}[1]{\nabla {#1}}

\usepackage{stmaryrd} 

\newcommand{\avg}[1]{\{\!\{#1\}\!\}}

\newcommand{\jump}[1]{\llbracket {#1} \rrbracket }
\newcommand{\jumporiented}[1]{\left[ {#1} \right] }

\newcommand{\intele}[2]{ \left( {#1},{#2} \right)_{\Omega_{e}} }
\newcommand{\inteleface}[2]{ \left( {#1},{#2} \right)_{\partial\Omega_{e}} }
\newcommand{\intelefaceInterior}[2]{ \left( {#1},{#2} \right)_{\partial\Omega_{e}\setminus\Gamma_h }}

\usepackage{enumitem}
\setlist[enumerate]{label*=\roman*),ref=\roman*)}

\journal{Journal}

\begin{document}

\begin{frontmatter}

\title{Efficiency of high-performance discontinuous Galerkin spectral element\\ methods for under-resolved turbulent incompressible flows}

\author{Niklas Fehn}
\ead{fehn@lnm.mw.tum.de}
\author{Wolfgang A. Wall}
\ead{wall@lnm.mw.tum.de}
\author{Martin Kronbichler\corref{correspondingauthor1}}
\cortext[correspondingauthor1]{Corresponding author at: Institute for Computational Mechanics, Technical University of Munich, Boltzmannstr. 15, 85748 Garching, Germany. Tel.: +49 89 28915300; fax: +49 89 28915301}
\ead{kronbichler@lnm.mw.tum.de}
\address{Institute for Computational Mechanics, Technical University of Munich,\\ Boltzmannstr. 15, 85748 Garching, Germany}

\begin{abstract}
The present paper addresses the numerical solution of turbulent flows with high-order discontinuous Galerkin methods for discretizing the incompressible Navier--Stokes equations. The efficiency of high-order methods when applied to under-resolved problems is an open issue in literature. This topic is carefully investigated in the present work by the example of the 3D Taylor--Green vortex problem. Our implementation is based on a generic high-performance framework for matrix-free evaluation of finite element operators with one of the best realizations currently known. We present a methodology to systematically analyze the efficiency of the incompressible Navier--Stokes solver for high polynomial degrees. Due to the absence of optimal rates of convergence in the under-resolved regime, our results reveal that demonstrating improved efficiency of high-order methods is a challenging task and that optimal computational complexity of solvers, preconditioners, and matrix-free implementations are necessary ingredients to achieve the goal of better solution quality at the same computational costs already for a geometrically simple problem such as the Taylor--Green vortex. Although the analysis is performed for a Cartesian geometry, our approach is generic and can be applied to arbitrary geometries. We present excellent performance numbers on modern, cache-based computer architectures achieving a throughput for operator evaluation of~$3\cdot 10^8$ up to~$1\cdot 10^9$ DoFs/sec on one Intel Haswell node with 28 cores. Compared to performance results published within the last 5 years for high-order DG discretizations of the compressible Navier--Stokes equations, our approach reduces computational costs by more than one order of magnitude for the same setup.
\end{abstract}

\begin{keyword}
Discontinuous Galerkin method, incompressible Navier--Stokes, implicit large-eddy simulation, high-performance computing, matrix-free implementation, high-order methods
\end{keyword}

\end{frontmatter}

\section{Introduction}\label{Intro}
The present paper deals with the numerical solution of the incompressible Navier--Stokes equations for turbulent flows using high-order discontinuous Galerkin methods for discretization in space and a state-of-the-art splitting method for discretization in time. Under certain assumptions like smoothness of the solution, high-order, spectral element methods are way more accurate than their low-order counterparts for the same number of unknowns, see also~\cite{Hesthaven07,Karniadakis13}. However, while optimal rates of convergence can be observed for simple analytical test cases with smooth solutions, the situation is different for more complex applications, especially for highly under-resolved, turbulent flows. In this regime, the assumption of optimal rates of convergence is rather inappropriate. One might argue that high-order methods are still more accurate for such problems due to smaller error constants and better dispersive properties. However, a fundamental question remains unanswered~\cite{Wang2013}: Do high-order methods improve the overall efficiency of the computational approach?

The absence of exact reference solutions when dealing with large eddy simulation of turbulent flows often hinders the analysis of this question since the accuracy of a particular method is difficult to evaluate in case that only reference solutions of the same solution quality are available or statistical errors are also present. Apart from that, analyzing this question is complicated by the fact that a precise investigation of this issue involves several more aspects beyond spatial discretization properties. It includes aspects related to the temporal discretization approach, the algorithmic complexity and effectiveness of iterative solvers and preconditioners, as well as the computational efficiency of the basic computational operations of a PDE solver such as the evaluation of discretized (non)linear spatial derivative operators (denoted as matrix--vector products), preconditioners and multigrid smoothers, as well as vector update operations. Since theoretical performance models, e.g., based on operation counts, tend to become inaccurate due to complexities introduced above, we favor a purely numerical investigation of the most relevant aspects.

The present incompressible Navier--Stokes DG solver has been developed within a series of recent publications. It is based on the high-order dual splitting scheme~\cite{Karniadakis1991} for discretization in time which has first been used in~\cite{Hesthaven07} in combination with discontinuous Galerkin methods for discretization in space. Instabilities of this discretization approach occuring for small time step sizes have been discussed and solved in~\cite{Krank2017,Fehn2017} by a proper DG discretization of the velocity--pressure coupling terms. A stabilization of the discontinuous Galerkin discretization for under-resolved flows based on a consistent divergence penalty term and a consistent continuity penalty term have been developed in~\cite{Krank2017,Fehn18}, which renders this approach a highly attractive candidate for implicit large eddy simulation. For example, this approach has been applied to large scale computations of turbulent flows such as direct numerical simulation of periodic hill flow in~\cite{Krank17b}. Our implementation is based on high-performance, matrix-free methods for tensor product finite elements developed recently in~\cite{Kronbichler2012, Kronbichler2017b}, exhibiting excellent performance characteristics~\cite{Kronbichler2017b,Kronbichler2017a}.

The outline of this paper is as follows. In Section~\ref{Efficiency}, we introduce a methodology for the evaluation of the efficiency of PDE solvers and define the relevant quantities in order to base the analysis on rational grounds. In Section~\ref{NumericalDiscretization}, the temporal discretization and the spatial discretization of the incompressible Navier--Stokes equations for the solution of under-resolved turbulent flows are briefly summarized. Numerical results are presented in Section~\ref{NumericalResults}. By focusing on the 3D Taylor--Green vortex problem, we present a detailed numerical analysis of the relevant quantities and factors determining the overall efficiency of the incompressible Navier--Stokes solver. In Section~\ref{Summary}, we discuss our results and give an outlook on future work.

\section{Efficiency of high-order discretizations for PDEs}\label{Efficiency}
\subsection{A general efficiency model for method-of-lines approaches and iterative solution techniques}
We consider method-of-lines type discretizations of partial differential equations where a system of (non)linear equations has to be solved within each time step by using iterative solution techniques. We want to acknowledge the fact that this is indeed an interdisciplinary effort that needs to bring together various ingredients from discretizations, numerical linear algebra, and computer science. In accordance with~\cite{Wang2013}, we define the overall efficiency~$E$ of a numerical method for the solution of a PDE as
\begin{align}
E = \frac{\text{accuracy}}{\text{computational\; costs}} \; ,\label{Efficiency1}
\end{align}
where accuracy is the inverse of a suitable and problem specific error measure and computational costs are the amount of computational resources measured as the wall time~$t_{\text{wall}}$ times the number of cores~$N_{\mathrm{cores}}$ in~\unit{CPUh}. In order to identify the contributions of the various disciplines, the above equation can be expanded highlighting the main components that determine the overall efficiency
\begin{align}
\begin{split}
E(h,k,\Delta t)&= \underbrace{\frac{\text{accuracy}}{\text{DoFs}\cdot\text{timesteps}}}_{\text{discretization}}
\cdot
\underbrace{\frac{1}{\text{iterations}}}_{\text{solvers/preconditioners}}
\cdot
\underbrace{\frac{\text{DoFs}\cdot\text{timesteps}\cdot\text{iterations}}{\text{computational costs}}}_{\text{implementation}} \\
&= E_{h,k,\Delta t}(h,k,\Delta t) \cdot E_{\bm{x}=\bm{A}^{-1}\bm{b}}(h,k,\Delta t) \cdot E_{\bm{A}\bm{x}}(h,k) \; .
\end{split}\label{Efficiency2}
\end{align}
The efficiency of the spatial and temporal discretization is denoted as~$E_{h,k,\Delta t}$ and essentially depends on the characteristic element length~$h$, the polynomial degree~$k$ of the shape functions, and the time step size~$\Delta t$. The order~$J$ of the time integration scheme does not appear explicitly as a parameter since it is considered to be constant in the present work; typically~$J=2$ is used for the type of incompressible Navier--Stokes solvers discussed in this work. The fact that the efficiency of the solvers and preconditioners and the efficiency of the implementation are written as seperate factors is based on the assumption that the application of discrete finite element operators is the basic ingredient of iterative solution techniques which have to be used for the large linear systems of equations arising from finite element discretizations of LES and DNS of turbulent flows. For the solution of linear systems of equations we use state-of-the-art iterative solution techniques such as the preconditioned conjugate gradient method. Hence, the efficiency~$ E_{\bm{x}=\bm{A}^{-1}\bm{b}}$ mainly depends on the type of preconditioner used to solve these equations. Note that iterations should not be understood as the number of iterations required to solve the linear system of equations but rather as an effective number of matrix--vector products (operator evaluations) applied during the whole iterative solution procedure including preconditioning operations. It also includes other operations such as level 1 BLAS vector operations (addition, scaling, inner product) which can account for a significant part of the overall computational costs. We do not define this quantity explicitly since we believe that this aspect is difficult to investigate theoretically, e.g., in terms of operation counts, but that it should be investigated numerically. To efficiently evaluate discretized finite element operators for high polynomial degrees we use a high-performance implementation based on matrix-free operator evaluations exploiting sum-factorization on tensor product elements. The efficiency of the implementation~$E_{\bm{A}\bm{x}}$ is defined as the number of degrees of freedom per time step and per operator evaluation that can be processed within a given amount of computational costs and is also denoted as throughput. It should be noted that the metric GFlops/sec often used to measure the efficiency of implementations is less relevant, i.e., performance optimizations should always target an increase in throughput rather than floating point operations. We emphasize that all components of the solver and preconditioner are implemented in a matrix-free way for our approach, which is crucial in order to obtain a method that is efficient as a whole. Hence, the efficiency of the implementation and the solver are highly connected for the spectral DG discretization methods considered in this work. Moreover, the condition number and hence the efficiency of the solvers and preconditioners~$ E_{\bm{x}=\bm{A}^{-1}\bm{b}}=E_{\bm{x}=\bm{A}^{-1}\bm{b}}(h,k,\Delta t)$ depends on the parameters of the spatial and temporal discretization. In general, the efficiency of the implementation~$E_{\bm{A}\bm{x}}=E_{\bm{A}\bm{x}}(h,k)$ also depends on the parameters of the spatial discretization. 

In previous works, the efficiency has been evaluated from different perspectives by concentrating on one of the three influence factors. For example, the accuracy of high-order DG discretizations has been analyzed in detail in~\cite{Fehn2017} for laminar flow problems and in~\cite{Fehn18} for turbulent flow problems. In the context of discontinuous Galerkin discretizations of the compressible Navier--Stokes equations, the accuracy of high-order methods has been investigated, e.g., in~\cite{Gassner2013,Wiart14} for the Taylor--Green vortex problem. Fast matrix-free methods for DG discretizations have been developed in~\cite{Kronbichler2017b,Muething2017} focusing on an efficient implementation that is optimized for modern cache-based computer architectures. Robust multigrid methods for high-order DG discretizations for elliptic and advection--diffusion problems have been developed in~\cite{Gopalakrishnan2003} and for adaptively refined grids in~\cite{Kanschat2004,Kanschat2008} using block Jacobi and block Gauss--Seidel smoothers. Robust~$p$-multigrid methods with overlapping Schwarz smoothers for high aspect ratios are analyzed in~\cite{Stiller2016}. In these works, the multigrid preconditioners are mainly discussed in terms of convergence rates and iteration counts. In the present work, multigrid smoothing is based on polynomial smoothers analyzed, e.g., in~\cite{Adams03,Sundar15} and used recently in~\cite{Krank2017,Carr2016} in the context of discontinuous Galerkin methods. Efficient matrix-free multigrid solvers taking into account the efficiency of both multigrid smoothers and the matrix-free implementation have been analyzed in~\cite{Kronbichler2016b} where the authors point to the importance of considering time-to-solution rather than iteration counts.

The present work fosters a more holistic view that focuses on the overall efficiency by considering  all relevant components. The analysis is dedicated to under-resolved turbulent flows using a high-order DG discretization. As explained above, exact reference resolutions are crucial for an objective quantification of errors (and hence the overall efficiency of the method) in the context of large eddy simulation. For this reason, we consider a physically complex problem on a simple geometry with well-defined initial conditions, the 3D Taylor--Green vortex problem, in favor of a detailed analysis of the relevant parameters~$h,k,\Delta t$ on the efficiency of the numerical method. We compute an accurate reference solution with a spatial resolution fine enough to resolve the relevant flow structures. The accuracy for under-resolved simulations can then be easily quantified by measuring the error of characteristic quantities such as the temporal evolution of the kinetic energy as compared to the exact reference solution. Of course, applying this analysis to geometrically complex, engineering applications is desirable, but we believe that central aspects of the performance of high-order DG methods can already be analyzed for the idealized case of a simple geometry. As pointed out in~\cite{Wang2013}, representing complex geometries needs careful design of suitable meshes which is beyond the scope of this work. However, it should be noted that our approach is generic and can be applied to arbitrary geometries.

\subsection{Assumptions and simplified efficiency model}
In the present paper, we will demonstrate that the proposed Navier--Stokes solver achieves optimal performance with respect to the implementation and to the solvers/preconditioners. For our approach, the number of degrees of freedom per time step and per iteration which can be processed within a given amount of computational costs is almost independent of the mesh refinement level and only mildly depends on the polynomial degree of the shape functions
\begin{align}
E_{\bm{A}\bm{x}}(h,k) \approx C(k) \approx \text{const} \; .\label{EfficiencySolvers}
\end{align}
This is a characteristic property of the matrix-free implementation using sum-factorization and is discussed in detail in this work. All systems of equations to be solved are symmetric, positive definite for which optimal complexity preconditioning strategies exist, i.e., the number of operator evaluations applied during the solution procedure is independent of the mesh refinement level. For the geometric multigrid methods with matrix-free, polynomial smoothing used in this work, the number of iterations mildly depends on the polynomial degree of the shape functions so that
\begin{align}
 E_{\bm{x}=\bm{A}^{-1}\bm{b}}(h,k,\Delta t)  \approx C(k) \approx \text{const}\; .\label{EfficiencyImplementation}
\end{align}
For operators including the mass matrix operator scaled by the inverse of the time step size, the linear system of equations might be solved more efficiently in the context of high-order, matrix-free DG methods by using the inverse mass matrix as a preconditioner instead of geometric multigrid preconditioners. In that case, the efficiency also mildly depends on~$h$ and~$\Delta t$, but the absolute efficiency is improved as compared to more complex preconditioners. We demonstrate that this preconditioner also allows to obtain mesh-independent convergence in case that the time step size is selected according to the CFL condition, i.e., the time step size is reduced under mesh refinement.

With the above assumptions, the spatial and temporal discretization emerges as the central impact factor for the overall efficiency of the numerical method. Regarding the temporal discretization, the time step size should be chosen so that temporal and spatial errors are comparable in order to maximize the overall efficiency. In case of a fully implicit Navier--Stokes solver, one typically applies this strategy and selects the time step size according to accuracy considerations only. In the present work, however, the analysis is restricted to an explicit treatment of the convective term while the viscous term is treated implicitly. As a result, the time step size is restricted according to the CFL (Courant--Friedrichs--Lewy) condition
\begin{align}
\Delta t = \frac{\mathrm{Cr}}{k_u^{r}}\frac{h_{\mathrm{min}}}{\Vert \bm {u} \Vert_{\mathrm{max}}} \; ,\label{CFL_Condition}
\end{align}
where~$\mathrm{Cr}$ denotes the Courant number,~$h_{\mathrm{min}}$ a characteristic length scale of the mesh calculated as the minimum vertex distance, and~$\Vert \bm {u} \Vert_{\mathrm{max}}$ the maximum velocity. The factor~$1/k_u^{r}$ highlights that the time step size has to be reduced for increasing polynomial degrees, where the exponent~$r$ of the polynomial degree~$k_u$ of the discrete velocity solution is discussed in more detail below. We will demonstrate numerically that an exponent of~$r=1.5$ accurately models the dependency on the polynomial degree, which has for example been used in~\cite{Fehn18}, as opposed to exponents of~$r=1$ or~$r=2$ used in other literature, see also~\cite{Hesthaven07}. For high Reynolds number flow problems with high spatial resolution requirements as considered in this work, one typically observes that the CFL condition is restrictive in the sense that the temporal discretization error is negligible for Courant numbers close to the critical value. Under this assumption, the overall efficiency is optimized by choosing the time step size as large as possible. As a result, the parameter~$\Delta t$ can be expressed as a function of the parameters~$h$ and~$k$ characterizing the spatial resolution, leading to the simplification~$E_{h,k,\Delta t}(h,k,\Delta t) = E_{h,k,\Delta t}(h,k)$. Hence, assuming optimality of the solvers and the implementation, the efficiency of the spatial discretization approach can be identified as a central point influencing the overall efficiency. 

Similar to equation~\eqref{Efficiency2}, the following relation can be derived for the computational costs
\begin{align}
\text{computational costs} = \underbrace{\text{DoFs}\cdot\text{timesteps}}_{\text{discretization}}
\cdot
\underbrace{\text{iterations}}_{\text{solvers/preconditioners}}
\cdot
\underbrace{\frac{\text{computational costs}}{\text{DoFs}\cdot\text{timesteps}\cdot\text{iterations}}}_{\text{implementation}} \; .
\end{align}
Using the relation~$N_{\mathrm{DoFs}} = C(k) h^{-d}$ for the number of unknowns~$N_{\mathrm{DoFs}}$,~$N_{\Delta t}\sim \Delta t^{-1}$ with~$\Delta t = C(k)h$ according to equation~\eqref{CFL_Condition} for the number of timesteps~$N_{\Delta t}$, and inserting the above assumptions (regarding the solution of linear systems of equations and the implementation), we obtain for the computational costs~$t_{\text{wall}}N_{\mathrm{cores}}$
\begin{align}
t_{\text{wall}}N_{\mathrm{cores}} = C(k) h^{-(d+1)} \; ,
\label{ComputationalCosts}
\end{align}
where the factor~$C(k)$ contains contributions from equations~\eqref{EfficiencySolvers},~\eqref{EfficiencyImplementation},~\eqref{CFL_Condition}, and the above relation for the number of unknowns~$N_{\mathrm{DoFs}}$.
\begin{figure}[!ht]
 \centering 
 \subfigure[Efficiency~$E_{h,k}$ of high-order spatial DG discretization in terms of error versus number of unknowns.]{
	 \includegraphics[width=0.8\textwidth]{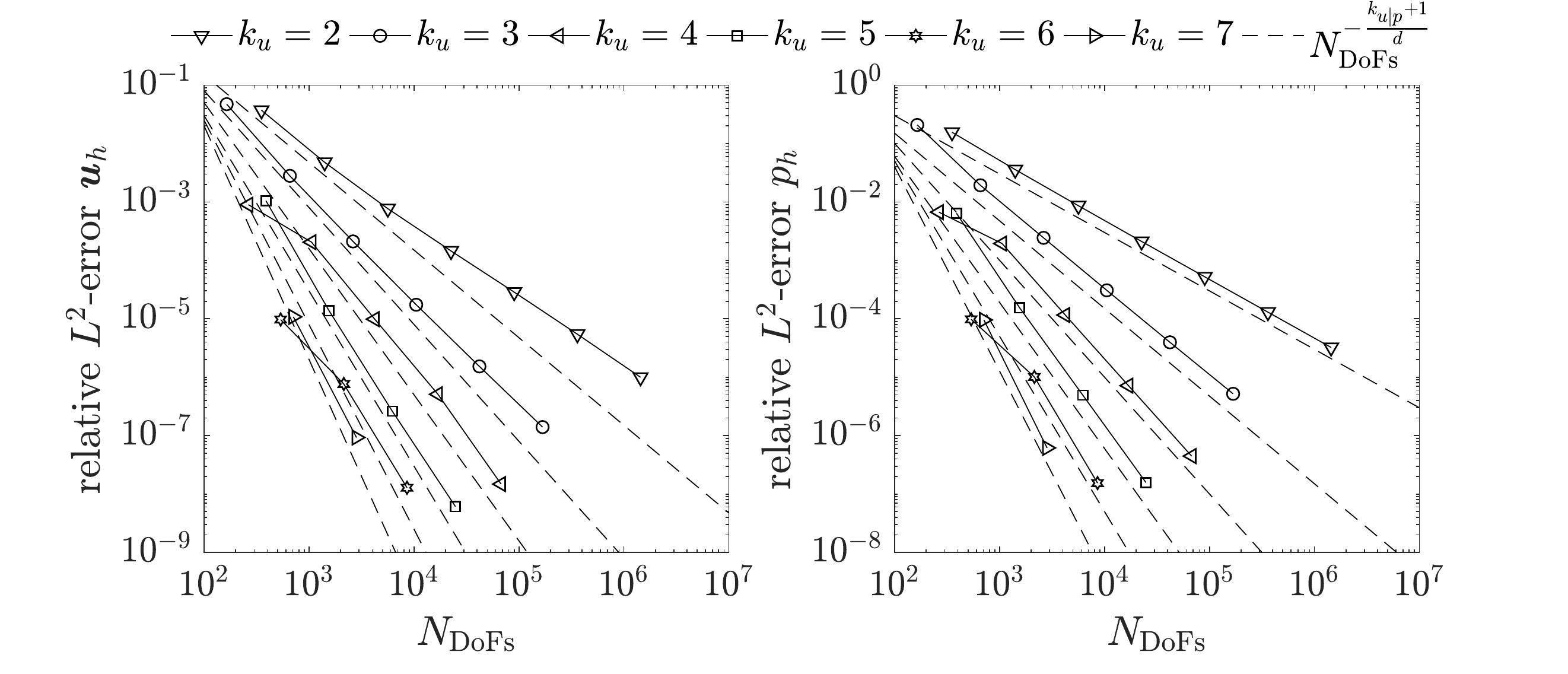}}
 \subfigure[Overall efficiency~$E$ in terms of error versus computational costs.]{
	 \includegraphics[width=0.8\textwidth]{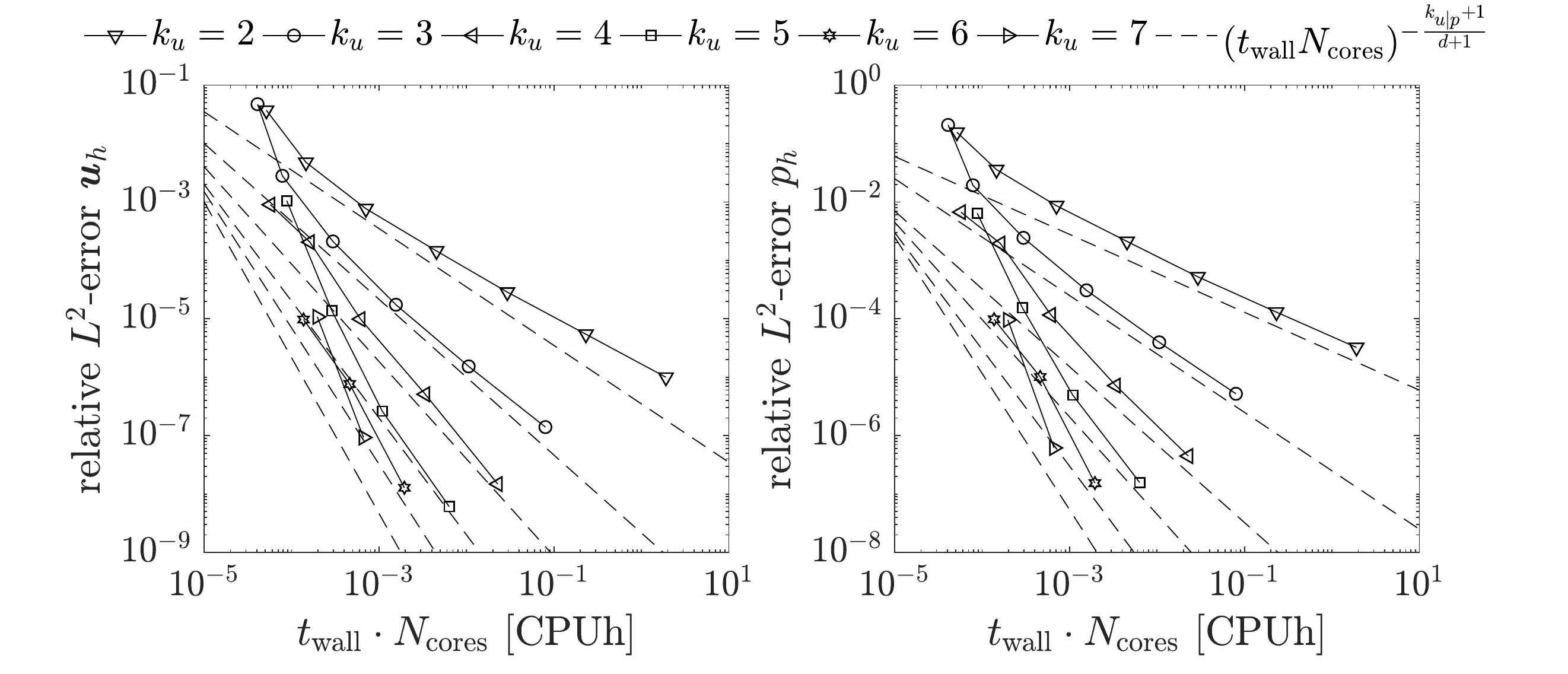}}
\caption{Two-dimensional vortex problem: $h$-convergence test for polynomial degrees~$k=2,...,7$ as well as evaluation of overall efficiency of high-order methods.}
\label{fig:2D_Vortex_Problem}
\end{figure}
We now distinguish two characteristic cases:
\begin{itemize}
\item The method shows optimal rates of convergence in space, i.e., the error is~$e=C(k) h^{k+1}$ (or equivalently~$e=C(k) N_{\mathrm{DoFs}}^{-\frac{k+1}{d}}$ in terms of number of unknowns) as typically observed for simple analytical test cases with smooth solutions. Using equation~\eqref{ComputationalCosts} we obtain
\begin{align}
e = C(k) \left(t_{\text{wall}}N_{\mathrm{cores}}\right)^{-\frac{k+1}{d+1}} \; .
\end{align}
When increasing the computational costs, the error decreases faster for increasing polynomial degrees. Consequently, one can expect that the efficiency of high-order methods with respect to the spatial discretization directly translates into an improved overall efficiency as compared to lower polynomial degrees. Such a situation is illustrated in Figure~\ref{fig:2D_Vortex_Problem} where results are shown for the two-dimensional vortex problem analyzed in~\cite{Krank2017,Fehn2017} in the context of high-order DG methods. We consider polynomial degrees~$k=2,...,7$, a viscosity of~$\nu=0.01$, and the BDF3 time integration scheme with~$\mathrm{Cr}=0.125$. The reader is referred to Section~\ref{NumericalResults}  for more details on how to perform these numerical experiments.

\item For under-resolved turbulent flows, one will typically not observe optimal rates of convergence in space but rather~$e = C(k) h^{\text{const}}$, i.e., the efficiency of high-order methods essentially depends on the unknown factor~$C(k)$. The constant exponent~$h^{\mathrm{const}}$ expresses that methods of different polynomial degrees show similar rates of convergence for comparable spatial resolutions ($N_{\mathrm{DoFs}}$), but it depends on the spatial resolution and eventually tends to the asymptotic convergence rates. In terms of computational costs we then obtain
\begin{align}
e = C(k) \left(t_{\text{wall}}N_{\mathrm{cores}}\right)^{-\frac{\text{const}}{d+1}} \; .
\end{align}
It is often stated in literature that high-order methods are superior for under-resolved situations in terms of accuracy, see for example~\cite{Gassner2013,Beck2014b}. However, it is unclear whether this also translates into superior efficiency of high-order methods, i.e., the factor~$C(k)$ is unknown and depends on several aspects as shown above. A numerical investigation of this aspect is subject of the present work.
\end{itemize}

\section{Numerical discretization of the incompressible Navier--Stokes equations}\label{NumericalDiscretization}
We seek numerical solutions of the incompressible Navier--Stokes equations in a domain~$\Omega \subset \mathbb{R}^d$
\begin{align}
\frac{\partial \bm{u}}{\partial t} + \nabla \cdot \bm{F}_{\mathrm{c}}(\bm{u}) - \nabla \cdot \bm{F}_{\mathrm{v}} (\bm{u}) + \nabla p &= \bm{f} \;\; \text{in}\; \Omega \times [0, T] \; ,\label{MomentumEquation}\\
\nabla \cdot \bm{u} &= 0  \;\; \text{in}\; \Omega \times [0, T] \; ,\label{ContinuityEquation}
\end{align}
where~$\bm{u}$ is the unknown velocity,~$p$ the kinematic pressure, and~$\bm{f}$ the body force vector. We use the conservative formulation of the convective term,~$\bm{F}_{\mathrm{c}}(\bm{u}) = \bm{u}\otimes \bm{u}$, and the Laplace formulation of the viscous term,~$\bm{F}_{\mathrm{v}} (\bm{u}) = \nu \Grad{\bm{u}}$, where the constant kinematic viscosity is denoted as~$\nu$.
The domain boundary is denoted by~$\Gamma = \partial \Omega$ and consists of periodic boundaries for the Taylor--Green vortex example considered in this work.
At time~$t=0$, a divergence-free initial condition is prescribed for the velocity field,~$\bm{u}(\bm{x},t=0) = \bm{u}_0(\bm{x})$ in~$\Omega$.

\subsection{Temporal discretization: High-order dual splitting scheme}\label{TemporalDiscretization}
Projection methods~\cite{Guermond06} are good candidates for high-performance numerical solutions of the incompressible Navier--Stokes equations due to their algorithmic simplicity. Instead of solving a coupled system of equations of indefinite saddle-point type for the velocity and pressure unknowns, the application of projection methods results in easier-to-solve equations such as a Poisson equation for the pressure and a convection-diffusion equation or Helmholtz like equation for the velocity unknowns (depending on the temporal treatment of the convective term). We note that an explicit treatment of the convective term introduces a restriction of the time step size~$\Delta t$ according to the CFL condition as opposed to an implicit formulation. At the same time, however, this strategy avoids the solution of nonlinear systems of equations, for which it is often difficult to develop robust and efficient (cost-effective) preconditioners for the convection-dominated regime. In the present work, we consider the dual splitting projection scheme developed in~\cite{Karniadakis1991}, for which the solution of each time step is split into four substeps
\begin{align}
\frac{\gamma_0\hat{\bm{u}}-\sum_{i=0}^{J-1}\left(\alpha_i\bm{u}^{n-i}\right)}{\Delta t} &= 
- \sum_{i=0}^{J-1}\left(\beta_i \Div{\bm{F}_{\mathrm{c}}\left(\bm{u}^{n-i}\right)}\right)
+ \bm{f}\left(t_{n+1}\right)\; ,\label{DualSplitting_ConvectiveStep}\\
-\nabla^2 p^{n+1} &= -\frac{\gamma_0 }{\Delta t}\Div{\hat{\bm{u}}} \; ,\label{DualSplitting_PressureStep}\\
\hat{\hat{\bm{u}}} &= \hat{\bm{u}} - \frac{\Delta t}{\gamma_0} \Grad{p^{n+1}}\; ,\label{DualSplitting_ProjectionStep}\\
\frac{\gamma_0 }{\Delta t} \bm{u}^{n+1}  -  \Div{\bm{F}_{\mathrm{v}}\left(\bm{u}^{n+1}\right)} &=
\frac{\gamma_0 }{\Delta t}\hat{\hat{\bm{u}}} \; .\label{DualSplitting_ViscousStep}
\end{align}
This splitting scheme is based on BDF time integration of order~$J$ with coefficients~$\gamma_0$,~$\alpha_i$,~$i=0,...,J-1$. The convective term is treated explicitly using an extrapolation scheme of order~$J$ with coefficients~$\beta_i$,~$i=0,...,J-1$ and is treated in the first substep of the splitting scheme along with the body force term. In the second substep, the pressure~$p^{n+1}$ is calculated by solving a Poisson equation where the divergence of the intermediate velocity field forms the right-hand side. Subsequently, the intermediate velocity~$\hat{\bm{u}}$ is projected onto the space of divergence-free vectors resulting in a second intermediate velocity field~$\hat{\hat{\bm{u}}}$. The viscous term is taken into account in the last substep resulting in the final velocity~$\bm{u}^{n+1}$. For the simulations performed in Section~\ref{NumericalResults}, the BDF2 scheme is used.

\subsection{Spatial discretization: High-order discontinuous Galerkin discretization}\label{SpatialDiscretization}
The spatial discretization is based on high-order discontinuous Galerkin methods. The computational domain~$\Omega_h = \bigcup_{e=1}^{N_{\text{el}}} \Omega_{e}$ consists of~$N_{\text{el}}$ non-overlapping quadrilateral/hexahedral elements. The numerical solution is discontinuous between elements and is approximated by polynomials of tensor degree~$\leq k$ within each element. Mixed-order polynomials of degree~$(k_u,k_p)=(k,k-1)$ for velocity and pressure are used in the present work for reasons of inf--sup stability~\cite{Fehn2017}, so that the spaces of test and trial functions for the velocity~${\bm{u}_h(\bm{x},t)\in\mathcal{V}^{u}_h}$ and the pressure~$p_h(\bm{x},t)\in \mathcal{V}^{p}_h$ are given as
\begin{align}
\mathcal{V}^{u}_{h} &= \left\lbrace\bm{u}_h\in \left[L_2(\Omega_h)\right]^d\; : \; \bm{u}_h\left(\bm{x}(\boldsymbol{\xi})\right)\vert_{\Omega_{e}}= \tilde{\bm{u}}_h^e(\boldsymbol{\xi})\vert_{\tilde{\Omega}_{e}}\in \mathcal{V}^{u}_{h,e}=[\mathcal{P}_{k_u}(\tilde{\Omega}_{e})]^d\; ,\;\; \forall e=1,\ldots,N_{\text{el}} \right\rbrace\;\; ,\\
\mathcal{V}^{p}_{h} &= \left\lbrace p_h\in L_2(\Omega_h)\; : \; p_h\left(\bm{x}(\boldsymbol{\xi})\right)\vert_{\Omega_{e}} = \tilde{p}_h^e(\boldsymbol{\xi})\vert_{\tilde{\Omega}_{e}}\in \mathcal{V}^{p}_{h,e}=\mathcal{P}_{k_p}(\tilde{\Omega}_{e})\; ,\;\; \forall e=1,\ldots,N_{\text{el}} \right\rbrace\; ,
\end{align}
Although we consider Cartesian geometries in the present work, the presented methods are generic in the sense that they extend naturally to complex geometries using a polynomial mapping of degree~$k_{\mathrm{m}}$,~$\bm{x}(\boldsymbol{\xi}) : \tilde{\Omega}_e =[0,1]^d\rightarrow \Omega_e$, for the transformation of the geometry from reference space to physical space. For the Cartesian meshes considered in this work, a linear mapping with degree~$k_{\mathrm{m}}=1$ is used. Multidimensional shape functions are given as the tensor product of one-dimensional shape functions where we use a nodal approach based on Lagrange polynomials with the Legendre--Gauss--Lobatto nodes as support points. As usual, volume and face integrals are written in shorthand notation as~$\intele{v}{u} = \int_{\Omega_e} v \odot u \; \mathrm{d}\Omega$ and~$\inteleface{v}{u} = \int_{\partial \Omega_e} v \odot u \; \mathrm{d} \Gamma$. The average operator~$\avg{\cdot}$ and the jump operators~$\jump{\cdot}$ and~$\jumporiented{\cdot}$ needed to define numerical fluxes on element faces are given as~$\avg{u} = (u^- + u^+)/2$,~$ \jump{u} = u^- \otimes \bm{n}^- + u^+ \otimes \bm{n}^+$, and~$\jumporiented{u}=u^- - u^+$, where~$(\cdot)^-$ denotes interior information,~$(\cdot)^+$ exterior information from the neighboring element, and~$\bm{n}$ the outward pointing unit normal vector. Volume and surface integrals occurring in the weak formulation are calculated using Gaussian quadrature. The number of one-dimensional quadrature points is chosen such that all integrals are calculated exactly in case of affine element geometries. We explicitly note that the increased computational costs associated with ``overintegration'' of the convective term according to the~$3/2$-rule is less of a concern since the convective term has to be evaluated only once within each time step. This aspect is, however, performance critical for high-order DG discretizations of the compressible Navier--Stokes equations as discussed in~\cite{Gassner2013,Flad2016}.

The weak discontinuous Galerkin discretization of the incompressible Navier--Stokes equations applied to the high-order dual splitting projection scheme can be summarized as follows: Find~$\hat{\bm{u}}_h,\hat{\hat{\bm{u}}}_h ,\bm{u}_h^{n+1}\in\mathcal{V}^u_h$ and~$p_h^{n+1}\in\mathcal{V}^p_h$ such that for all~$\bm{v}_h \in \mathcal{V}^{u}_{h,e}$,~$q_h \in \mathcal{V}^{p}_{h,e}$ and for all elements~$e=1,...,N_{\text{el}}$
\begin{align}
m^{e}_{h,u}\left(\bm{v}_h,\frac{\gamma_0 \hat{\bm{u}}_h-\sum_{i=0}^{J-1}\left(\alpha_i\bm{u}^{n-i}_h\right)}{\Delta t} \right)
&= 
- \sum_{i=0}^{J-1} \left(\beta_i c^e_h\left(\bm{v}_h,\bm{u}^{n-i}_h\right)\right)
+ \intele{\bm{v}_h}{\bm{f}(t_{n+1})} \; ,
\label{DualSplitting_ConvectiveStep_WeakForm}\\
l_{h}^{e}\left(q_h,p_h^{n+1}\right) &= - \frac{\gamma_0}{\Delta t} d_{h}^{e}\left(q_h,\hat{\bm{u}}_h\right)
\; ,
\label{DualSplitting_Pressure_WeakForm}\\
m_{h,u}^{e}(\bm{v}_h,\hat{\hat{\bm{u}}}_h)+ a^e_{\mathrm{D}}(\bm{v}_h,\hat{\hat{\bm{u}}}_h) + a^e_{\mathrm{C}}(\bm{v}_h,\hat{\hat{\bm{u}}}_h)  &= m_{h,u}^{e}\left(\bm{v}_h,\hat{\bm{u}}_h\right)-\frac{\Delta t}{\gamma_0}g_h^{e}\left(\bm{v}_h,p_h^{n+1}\right)\; ,\label{DualSplitting_Projection_WeakForm}\\
m^{e}_{h,u}\left(\bm{v}_h,\frac{\gamma_0}{\Delta t} \bm{u}_h^{n+1} \right) 
+ v^{e}_{h}\left(\bm{v}_h,\bm{u}_h^{n+1}\right)
&= 
m^{e}_{h,u}\left(\bm{v}_h,\frac{\gamma_0}{\Delta t}\hat{\hat{\bm{u}}}_h \right)
\; ,
\label{DualSplitting_ViscousStep_WeakForm}
\end{align}
where~$m^e_{h,u}\left(\bm{v}_h,\bm{u}_h\right)$ is the (velocity) mass matrix term,~$c^e_h\left(\bm{v}_h,\bm{u}_h\right)$ the convective term,~$v^e_h\left(\bm{v}_h,\bm{u}_h\right)$ the viscous term,~$g^e_h\left(\bm{v}_h,p_h\right)$ the pressure gradient term,~$d^e_h\left(q_h,\bm{u}_h\right)$ the velocity divergence term, and~$l_h^e\left(q_h,p_h\right)$ the (negative) Laplace operator. We use the local Lax--Friedrichs flux for the discretization of the convective term~\cite{Hesthaven07,Shahbazi07}. Central fluxes are used for the pressure gradient term and the velocity divergence term. Note that this is in contrast to the DG discretizations used previously for the dual splitting scheme in~\cite{Hesthaven07,Ferrer11} where no integration by parts is performed for the velocity--pressure coupling terms. Recently, it has been demonstrated in~\cite{Fehn2017} that the instabilities for small time steps and coarse spatial resolutions observed and discussed in~\cite{Ferrer11,Ferrer14} are related to the DG discretization of the velocity--pressure coupling terms and that integration by parts of these terms along with the definition of consistent boundary conditions for the intermediate velocity field~$\hat{\bm{u}}$ should be performed to obtain a stable discretization scheme. The viscous term as well as the negative Laplace operator in the pressure Poisson equation are discretized using the symmetric interior penalty method~\cite{Arnold2002}. Accordingly, the weak DG formulation of the above operators can be summarized as follows (see~\cite{Fehn2017,Fehn18} and references therein for more detailed derivations)
\begin{align}
m_{h}^{e}(\bm{v}_h,\bm{u}_h) &= \intele{\bm{v}_h}{\bm{u}_h}\; ,\\
c^e_h\left(\bm{v}_h,\bm{u}_h\right) &= -\intele{\Grad{\bm{v}_h}}{\bm{F}_{\mathrm{c}}(\bm{u}_h)} + 
\inteleface{\bm{v}_h}{\left(\avg{\bm{F}_{\mathrm{c}}(\bm{u}_h)} + \max \left(\vert \bm{u}_h^{-} \cdot \bm{n}\vert ,\vert \bm{u}_h^{+} \cdot \bm{n}\vert\right)\jump{\bm{u}_h}  \right) \cdot \bm{n}}\\
g^e_h\left(\bm{v}_h,p_h\right) &= -\intele{\Div{\bm{v}_h}}{p_h}+\inteleface{\bm{v}_h}{ \avg{p_h}\bm{n}}\; ,\\
d^e_h\left(q_h,\bm{u}_h\right) &= -\intele{\Grad{q_h}}{\bm{u}_h}+\inteleface{q_h}{\avg{\bm{u}_h}\cdot\bm{n}}\; ,\\
v_{h}^{e}(\bm{v}_h,\bm{u}_h) &= 
 \intele{\Grad{\bm{v}_h}}{\nu \Grad{\bm{u}_h}}
  - \inteleface{\Grad{\bm{v}_h}}{\nu/2\; \jump{\bm{u}_h}}
  - \inteleface{\bm{v}_h}{\nu \avg{\Grad{\bm{u}_h}}\cdot\bm{n}}
  + \inteleface{\bm{v}_h}{\nu\tau \jump{\bm{u}_h}\cdot\bm{n}}\; ,\\
  l_h^e\left(q_h,p_h\right) &= \intele{\Grad{q_h}}{\Grad{p_h}}
-\inteleface{\Grad{q_h}}{1/2\;\jump{p_h}}
- \inteleface{q_h}{\avg{\Grad{p_h}}\cdot\bm{n}}
+ \inteleface{q_h}{\tau\jump{p_h}\cdot\bm{n}}	\; .
\end{align}
In order to stabilize the DG discretization for convection-dominated, under-resolved flows, a consistent divergence penalty term~$a^e_{\mathrm{D}}(\bm{v}_h,\bm{u}_h)$  and a consistent continuity penalty term $a^e_{\mathrm{C}}(\bm{v}_h,\bm{u}_h)$ weakly enforcing the incompressibility constraint and inter-element continuity of the velocity, respectively, are added to the weak formulation of the projection step~\cite{Fehn18}
\begin{align}
a^e_{\mathrm{D}}(\bm{v}_h,\bm{u}_h) &= \intele{\Div{\bm{v}_h}}{\tau_{\mathrm{D}}\Div{\bm{u}_h}} \; \text{with} \;\;\tau_{\mathrm{D},e}=\zeta_{\mathrm{D}} \; \overline{\Vert\bm{u}^{n+1,\mathrm{ex}}_h \Vert} \; \frac{h}{k_u + 1} \; \Delta t \; ,\label{DivergencePenaltyTerm}\\
a^e_{\mathrm{C}}(\bm{v}_h,\bm{u}_h)&=\intelefaceInterior{\bm{v}_h\cdot \bm{n}}{\avg{\tau_{\mathrm{C},e}}\jumporiented{\bm{u}_h}\cdot \bm{n}} \; \text{with}  \;\; \tau_{\mathrm{C},e}=\zeta_{\mathrm{C}}\;\overline{\Vert \bm{u}^{n+1,\mathrm{ex}}_h \Vert} \; \Delta t \; ,\label{ContinuityPenaltyTerm}
\end{align}
where~$\overline{\Vert\bm{u}^{n+1,\mathrm{ex}}_h\Vert}=\overline{\Vert\sum_{i=0}^{J-1} \left(\beta_i \bm{u}^{n-i}_h\right)\Vert}$ is the elementwise, volume-averaged norm of the extrapolated velocity field and~$h/(k_u+1)=V_e^{1/3}/(k_u+1)$ an effective element length where~$V_e$ is the element volume. In the present work,~$\zeta_{\mathrm{D}}=\zeta_{\mathrm{C}}=1$ is used. Similar terms have first been proposed in~\cite{Steinmoeller13,Joshi16} as a means to stabilize the spatially discretized pressure projection operator and have been used in~\cite{Krank2017} in the context of large eddy simulation in combination with the dual splitting scheme. Recently, the divergence and continuity penalty terms have been anaylzed rigorously in~\cite{Fehn18}, where it has been demonstrated that these terms lead to a robust and accurate discretization scheme for under-resolved flows independently of the solution strategy used to solve the incompressible Navier--Stokes equations. These penalty terms can be motivated from the point of view of mass conservation as well as energy stability and might be interpreted as a realization of divergence-free~$H(\mathrm{div})$ elements in a weak sense.

\subsection{Solution of linear systems of equations and efficient matrix-free preconditioning}
Written in matrix notation, the four substeps of the dual splitting scheme are given as follows
\begin{align}
\bm{M}\frac{\gamma_0 \hat{\bm{U}}-\sum_{i=0}^{J-1}\left(\alpha_i\bm{U}^{n-i}\right)}{\Delta t}
&= 
- \sum_{i=0}^{J-1} \left(\beta_i \bm{C}\left(\bm{U}^{n-i}\right) \right)
+ \bm{F}(t_{n+1}) \; ,
\label{DualSplitting_ConvectiveStep_MatrixForm}\\
\bm{L}_{\text{hom}}\bm{P}^{n+1} &= - \frac{\gamma_0}{\Delta t} \bm{D}\hat{\bm{U}} - \bm{L}_{\text{inhom}}
\; ,
\label{DualSplitting_Pressure_MatrixForm}\\
\left(\bm{M} + \bm{A}_{\mathrm{D}} + \bm{A}_{\mathrm{C}} \right)\hat{\hat{\bm{U}}}  &= \bm{M}\hat{\bm{U}}-\frac{\Delta t}{\gamma_0}\bm{G}\bm{P}^{n+1}\; ,\label{DualSplitting_Projection_MatrixForm}\\
\left(\frac{\gamma_0}{\Delta t} \bm{M}
+ \bm{V}_{\mathrm{hom}}\right)\bm{U}^{n+1}
&= 
\frac{\gamma_0}{\Delta t}\bm{M}\hat{\hat{\bm{U}}}-\bm{V}_{\mathrm{inhom}}
\; ,
\label{DualSplitting_ViscousStep_MatrixForm}
\end{align}
To solve the convective step, the inverse mass matrix has to be applied to the right-hand side of equation~\eqref{DualSplitting_ConvectiveStep_MatrixForm}. In case of discontinuous Galerkin discretizations, the mass matrix is block-diagonal and can be inverted in an elementwise manner. More importantly, the inverse mass matrix operation can be implemented in a matrix-free way~\cite{Kronbichler2016} by using tensorial sum factorization with costs comparable to the forward application of the mass matrix. Hence, there is no need to choose the quadrature points equal to the interpolation points of the nodal shape functions. In Section~\ref{NumericalResults}, we demonstrate that the inverse mass matrix operation is a memory-bound operation and is as expensive as scaling a vector by a diagonal mass matrix.

The pressure Poisson equation~\eqref{DualSplitting_Pressure_MatrixForm}, the projection equation~\eqref{DualSplitting_Projection_MatrixForm}, and the Helmholtz equation~\eqref{DualSplitting_ViscousStep_MatrixForm} form symmetric, positive definite linear systems of equations. For the numerical solution of these equations, the preconditioned conjugate gradient (CG) method is used with an absolute solver tolerance of~$10^{-12}$ and a relative solver tolerance of~$10^{-6}$, where relative tolerance means that the residual is reduced by this factor compared to the initial guess that is computed as the extrapolation of the solution from previous time steps. As a preconditioner for the pressure Poisson equation, one V-cycle of the geometric multigrid method is used with a polynomial Chebyshev smoother~\cite{Adams03} resulting in mesh-independent convergence rates. The degree of the Chebyshev smoother is~$5$ for both pre- and postsmoothing and the smoothing range is~$20$ for all computations resulting in 13 matrix-vector products on the finest level per CG iteration. Note, however, that all multigrid operations are performed in single precision increasing the efficiency by a factor of approximately 2. As a coarse grid solver, we also use the Chebyshev method with a fixed number of iterations. The diagonal of the matrix required by the Chebyshev smoother is pre-computed in the setup phase. Accordingly, all components of the multigrid method including level transfer operators are implemented in a matrix-free way and we refer to~\cite{Krank2017} for more detailed information. For the projection equation as well as the Helmholtz equation, the inverse mass matrix is an effective preconditioner due to the relatively small time step sizes related to the CFL condition, see also~\cite{Shahbazi07}. Note that the operators~$\bm{A}_{\mathrm{D}}$ and~$\bm{A}_{\mathrm{C}}$ include the factor~$\Delta t$ according to equations~\eqref{DivergencePenaltyTerm} and~\eqref{ContinuityPenaltyTerm}, respectively. The mass matrix can be inverted very efficiently in a matrix-free way so that this preconditioner is highly efficient.

\subsection{Matrix-free evaluation of DG-discretized operators}

The algorithm is implemented in the \texttt{C++} programming language using the object-oriented finite element library~\texttt{deal.II}~\cite{dealII85}. Our implementation computes all matrix-vector products and right-hand side vectors by matrix-free operator evaluation based on fast integration, using the high-performance framework developed in~\cite{Kronbichler2012, Kronbichler2017b}. The integrals over cells and faces are computed by sum factorization specialized for quadrilateral and hexahedral cell shapes. These algorithms have their origin in the spectral element community~\cite{Orszag1980} and exploit the tensor product structure in the shape functions and the quadrature points. This allows to transform summations over~$(k+1)^d$ solution coefficients for each of the~$(k+1)^d$ quadrature points into~$d$ sums of length~$k+1$ for each point. Thus, evaluating all integrals pertaining to a cell involves~$\mathcal O((k+1)^{d+1})$ arithmetic operations. Likewise, face integrals are of cost~$\mathcal O((k+1)^{d})$. When expressed in terms of the number of degrees of freedom, the number of arithmetic operations per degree of freedom scales as~$\mathcal O(k)$ in the polynomial degree~$k$. At low and moderate polynomial degrees, the face integrals and the data access, which scale as~$\mathcal O(1)$ per degree of freedom, dominate over the cost of cell integrals~\cite{Kronbichler2017b}. Hence, the observed throughput of degrees of freedom processed per second appears to be almost independent of the polynomial degree for a wide range of polynomial degrees.

As the resulting kernel is relatively compute-heavy with around 100 to 400 arithmetic operations per degree of freedom for e.g.~the DG discretization of the Laplacian~\cite{Kronbichler2017b}, the minimal use of arithmetic instructions is essential for performance. Our realization utilizes an on-the-fly basis change between the given polynomial basis and a collocation basis of Lagrange polynomials in the points of the Gaussian quadrature~\cite{Kronbichler2017b}. With this technique, the number of tensor product calls for evaluating the cell Laplacian is~$4d=12$ in three spatial dimensions and~$2d=6$ for the mass matrix operator. Note that we do not use a collocation basis because that basis gives worse iteration numbers in multigrid algorithms with point-Jacobi smoothers than Lagrange polynomials in the Legendre--Gauss--Lobatto points. Also, Lagrange polynomials in Legendre--Gauss--Lobatto points enable cheaper face integrals as discussed in~\cite{Kronbichler2016}. Furthermore, operations in one-dimensional interpolation and derivative kernels of sum factorization are cut into half by using the even-odd decomposition~\cite{Kopriva09}.

With respect to computer architecture, our implementation explicitly uses 4-wide vector registers enabled by the AVX2 instruction set extension including fused multiply-add instructions on the Intel Haswell processors used for the experiments in this study. Vectorization is performed over several cells and faces, respectively, which gives best performance according to the analysis in~\cite[Sect.~3.2]{Kronbichler2017b}. Despite the full flexibility of DG methods with respect to geometry representation and nonlinear operators, our implementation is close to being memory bandwidth bound and thus almost reaching the throughput of simple finite difference stencils. Finally, all solver components have been completely parallelized with the message passing interface (MPI) and tuned to scale to~$\mathcal{O}(10^5)$ processors and tens of billions of unknowns as shown in~\cite{Krank2017}, in particular the iterative linear solvers, even though the computations performed in this work with up to~$4\cdot 10^9$ unknowns run on a smaller scale. In the field of high-order continuous finite element methods, strong scaling experiments up to~$5\cdot 10^5$ cores and approximately two billion grid points have been shown in~\cite{Offermans2016} for the spectral element solver Nek5000.

\section{Numerical results -- 3D Taylor--Green vortex problem}\label{NumericalResults}
As a numerical test case we consider the 3D Taylor--Green vortex problem~\cite{Taylor1937} which has been analyzed for example in~\cite{Fehn18,Gassner2013,Wiart14,Piatkowski16} in the context of high-order discontinuous Galerkin discretizations of the compressible and incompressible Navier--Stokes equations. This benchmark problem is defined by the following initial velocity field
\begin{align*}
u_1(\bm{x},t=0) &= U_0\sin\left(\frac{x_1}{L} \right)\cos\left(\frac{x_2}{L}\right)\cos\left(\frac{x_3}{L}\right)\; ,\\
u_2(\bm{x},t=0) &= -U_0\cos\left(\frac{x_1}{L} \right)\sin\left(\frac{x_2}{L}\right)\cos\left(\frac{x_3}{L}\right)\; ,\\
u_3(\bm{x},t=0) &= 0\; .
\end{align*}
The incompressible Navier--Stokes equations are solved on the computational domain~$\Omega_h = [-\pi L,\pi L]^3$ with length scale~$L$. The forcing term is zero,~$\bm{f}=\textbf{0}$, and periodic boundary conditions are prescribed in all directions. The initially smooth velocity field with large vortices breaks down into smaller eddies and a chaotic and turbulent flow develops, rendering this problem an interesting and challenging benchmark problem for the validation of turbulent flow solvers. In the present work, the Reynolds number is set to~$\mathrm{Re}=\frac{U_0 L}{\nu}=1600$ using the parameters~$U_0=1$,~$L=1$,~$\nu=\frac{1}{\mathrm{Re}}$. The time interval is~$0\leq t \leq T$ where the end time is given as~$T=20 T_0$ with~$T_0=\frac{L}{U_0}$, see also~\cite{Wang2013}. 

The computational domain is discretized using a uniform Cartesian grid consisting of~$(2^l)^d$ elements where~$l$ denotes the level of refinement. The number of unknowns is given as~$N_{\mathrm{DoFs}}=(2^l)^d(d(k_u+1)^d+(k_p+1)^d)=(2^l)^d(d(k+1)^d+k^d)$ and we define the effective spatial resolution as~$ (2^l(k+1))^d$, e.g., the effective resolution is~$64^3$ for refine level~$l=3$ with 8 elements per direction and polynomial degree~$k=7$. The polynomial degrees considered in the following are~$k=2,3,5,7,11,15$, where~$k=2$ serves as a reference low-order method in order to evaluate the efficiency for increasing polynomial degree~$k$. These polynomial degrees are selected since~$k=2,5,11$ and~$k=3,7,15$ allow to obtain the same effective resolution for different refinement levels, respectively. The time step size is calculated according to equation~\eqref{CFL_Condition} with~$\Vert \bm {u} \Vert_{\mathrm{max}}=U_0$ and~$h_{\mathrm{min}}=h$.


In the remainder of this chapter we analyze the performance of the present DG Navier--Stokes solver for the Taylor--Green vortex problem. As a prerequisite, Section~\ref{TGV_SelectionOfTimeStepSize} discusses how to select the time step size in an optimal way. Then, following the methodology introduced in Section~\ref{Efficiency}, we first analyze the efficiency of the spatial discretization and the temporal discretization in Section~\ref{TGV_EfficiencyDiscretization}, the efficiency of the solvers and preconditioners in Section~\ref{TGV_EfficiencySolvers}, and the efficiency of the matrix-free implementation in Section~\ref{TGV_EfficiencyImplementation}. Finally, the overall efficiency of the computational approach is discussed in Section~\ref{TGV_OverallEfficiency}.

\subsection{Selection of time step size and CFL condition}\label{TGV_SelectionOfTimeStepSize}
\begin{table}
\renewcommand{\arraystretch}{1.1}
\begin{center}
\begin{small}
\begin{tabular}{lllllll}
\hline
$\mathrm{Cr}$ number   & \multicolumn{6}{l}{Polynomial degree}\\
\cline{2-7}
					   & $k=2$ & $k=3$ & $k=5$ & $k=7$ & $k=11$ & $k=15$\\
\hline
$\mathrm{Cr}$ stable   & $0.20$ & $0.21$ & $0.22$ & $0.19$ & $0.22$ & $0.18$\\
$\mathrm{Cr}$ unstable & $0.21$ & $0.22$ & $0.23$ & $0.20$ & $0.23$ & $0.19$\\
\hline
\end{tabular}
\end{small}
\end{center}
\renewcommand{\arraystretch}{1}
\caption{Experimental determination of critical Courant number as a function of the polynomial degree~$k$ using~$r=1.5$. The refinement level is~$l=4$ for~$k=2,3$,~$l=3$ for~$k=5,7$, and~$l=2$ for~$k=11,15$.}
\label{CriticalCourantNumbers}
\end{table}

To evaluate the efficiency of the presented Navier--Stokes solver as a function of the polynomial degree, it is a necessary prerequisite to select the time step size in an optimal way. Due to the explicit formulation of the convective term, the time step size is restricted according to the CFL condition~\eqref{CFL_Condition}. Very generally, the time step size should be chosen so that temporal and spatial discretization errors are comparable for a simulation to be efficient. The CFL condition, however, is restrictive in the sense that the overall error is typically dominated by the spatial discretization error for~$\mathrm{Cr}<\mathrm{Cr}_{\mathrm{crit}}$ and for under-resolved problems as considered in this work. Hence, we select the time step size close to the critical time step size arising from the CFL condition in order to obtain the best efficiency for every spatial resolution. Therefore, one first has to verify the correctness of relation~\eqref{CFL_Condition} and the exponent~$r$ to allow a fair comparison of DG discretizations of different polynomial degrees. This is done experimentally by simulating the Taylor--Green vortex problem for different polynomial degrees and varying the Courant number to obtain a lower as well as an upper bound for the critical Courant number. For these investigations, we use refinement level~$l=4$ for~$k=2,3$, refinement level~$l=3$ for~$k=5,7$, and refinement level~$l=2$ for~$k=11,15$. The results in Table~\ref{CriticalCourantNumbers} demonstrate that an exponent of~$r=1.5$ in equation~\eqref{CFL_Condition} perfectly models the critical Courant number as a function of the polynomial degree for the Taylor--Green vortex problem considered here. Consequently, we can select a constant~$\mathrm{Cr}$ number for all polynomial degrees for the following calculations ensuring that neither low polynomial degrees nor high polynomial degrees benefit from the selection of the time step size in terms of the efficiency calculations detailed below. It turned out that the critical~$\mathrm{Cr}$ number is slightly decreasing for fine spatial resolutions. Hence, we use~$\mathrm{Cr} = 0.125$ to obtain stability for all spatial resolutions simulated in the following. This also implies that the overall runtime could be reduced for most simulations  with coarser spatial resolution. However, this procedure is acceptable since we are mainly interested in the performance as a function of the polynomial degree.

\subsection{Efficiency of discretization scheme}\label{TGV_EfficiencyDiscretization}

\begin{figure}[!ht]
 \centering 
 \subfigure[Effective resolution of~$64^3$ for different combinations of refinement level~$l$ and polynomial degree~$k$.]{
	\includegraphics[width=1.0\textwidth]{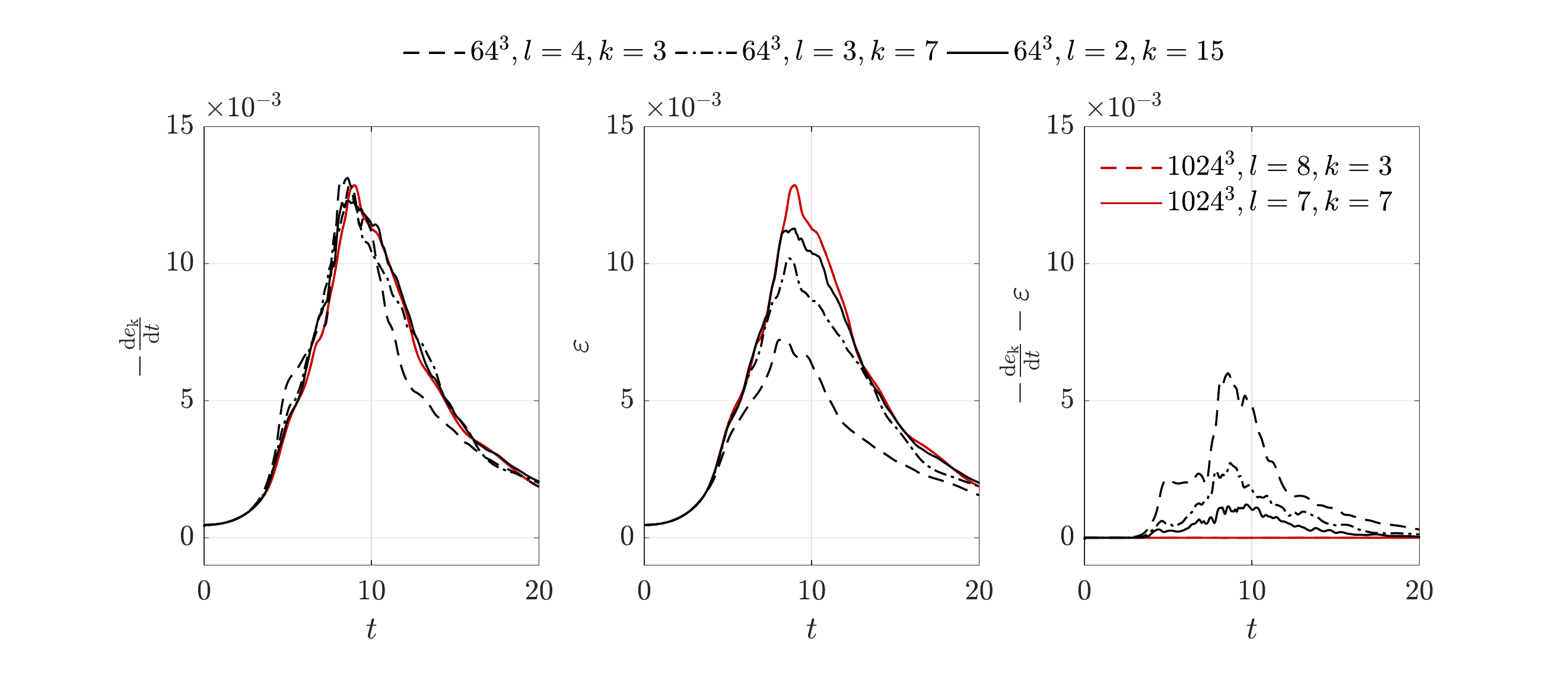}}
 \subfigure[Effective resolution of~$128^3$ for different combinations of refinement level~$l$ and polynomial degree~$k$.]{
	\includegraphics[width=1.0\textwidth]{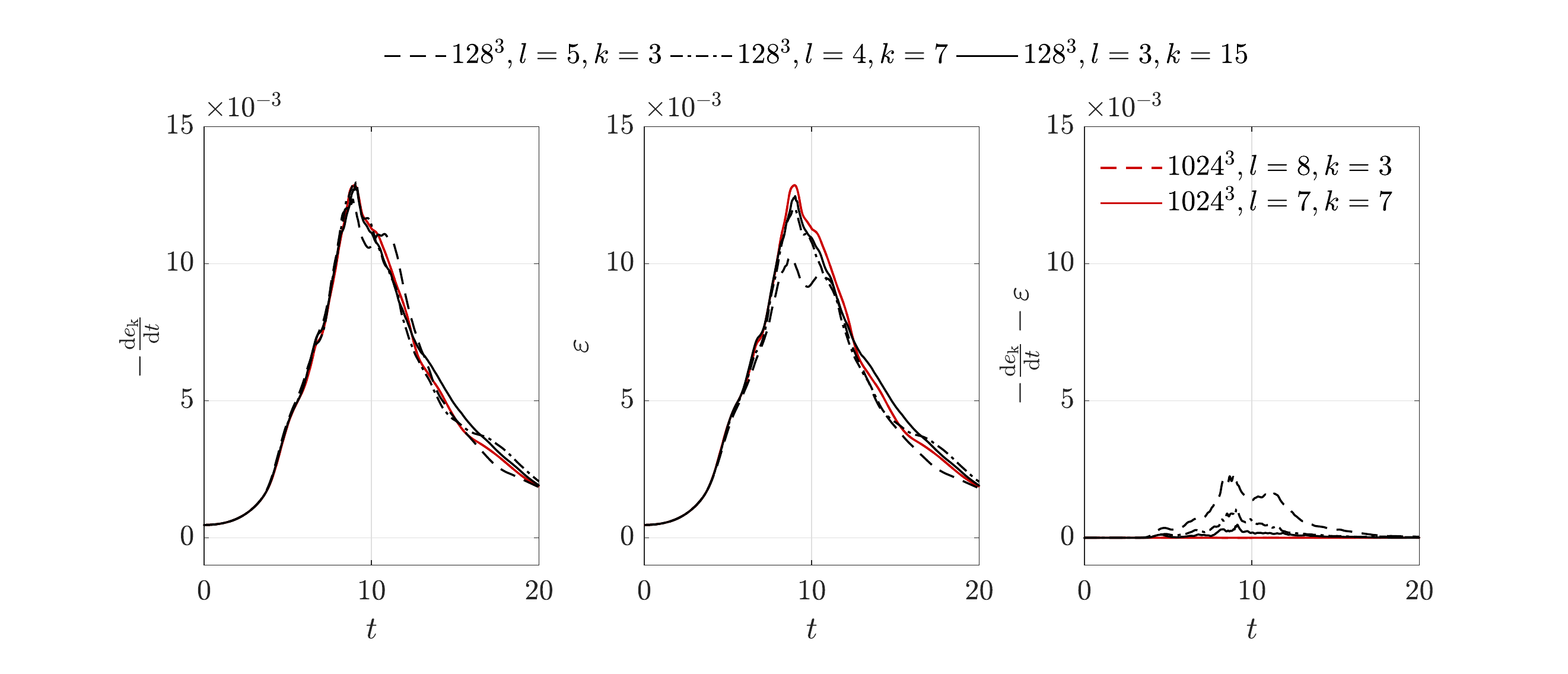}}
\caption{Rate of change of kinetic energy, molecular dissipation, and numerical dissipation as a function of time using effective resolutions of~$64^3$ and~$128^3$ velocity degrees of freedom for polynomial degrees~$k=3,7,15$. The results are compared to accurate reference solutions computed on meshes with an effective resolution of~$1024^3$ for polynomial degrees~$k=3,7$.}
\label{fig:3D_Taylor_Green_Kinetic_Energy_Dissipation}
\end{figure}

In order to evaluate the efficiency of the spatial discretization scheme w.r.t.~quality of the solution for a given number of unknowns we analyze the temporal evolution of the kinetic energy, the kinetic energy dissipation rate, and the numerical dissipation, defined as
\begin{align*}
e_{\mathrm{k}} = \frac{1}{V_{\Omega_h}}\int_{\Omega_h} \frac{1}{2} \bm{u}_h\cdot \bm{u}_h\; \mathrm{d}\Omega \; ,\hspace{0,5cm}
\varepsilon = \frac{\nu}{V_{\Omega_h}} \int_{\Omega_h} \Grad{\bm{u}_h} : \Grad{\bm{u}_h}\; \mathrm{d}\Omega \; ,\hspace{0,5cm}
\varepsilon_{\mathrm{num}} = -\frac{\mathrm{d}e_{\mathrm{k}}}{\mathrm{d}t}-\varepsilon \; .
\end{align*}
The numerical dissipation~$\varepsilon_{\mathrm{num}}$ is the difference between the time derivative of the kinetic energy and the dissipation rate of the resolved scales, and therefore quantifies the dissipation realized by the discretization scheme. For more detailed information on how to compute these quantities numerically the reader is referred to~\cite{Fehn18}. In Figure~\ref{fig:3D_Taylor_Green_Kinetic_Energy_Dissipation}, results are presented considering effective spatial resolutions of~$64^3$ and~$128^3$ for three different polynomial degrees~$k=3,7,15$. Accurate solutions obtained on meshes with an effective resolution of~$1024^3$ for polynomial degrees~$k=3,7$ serve as reference solution. For increasing spatial resolution, the results converge towards the reference solution. Note that it has been shown in~\cite{Fehn18} that the present discretization approach is competitive in accuracy to state-of-the-art implicit LES models in the context of finite volume methods or high-order DG discretizations of the compressible Navier--Stokes equations. On the one hand, we observe that the numerical dissipation is continuously decreasing for increasing polynomial degree~$k$ for a fixed spatial resolution. One the other hand, the approximation of the dissipation rate~$\varepsilon$ of the resolved scales is clearly improved for higher polynomial degrees. Although the highest polynomial degree~$k=15$ appears to be the most accurate one in terms of the kinetic energy evolution~$\frac{\mathrm{d}e_{\mathrm{k}}}{\mathrm{d}t}$, both effects seem to counterbalance each other so that the differences between low and high-order methods are less obvious for the temporal evolution of the overall kinetic energy~$e_{\mathrm{k}}$ as compared to the dissipation rate~$\varepsilon$, see also the discussion in~\cite{Fehn18}.

\begin{figure}[!ht]
 \centering 
 \subfigure[Efficiency~$E_{h,k}$ of spatial discretization approach.]{
	 \includegraphics[width=1.0\textwidth]{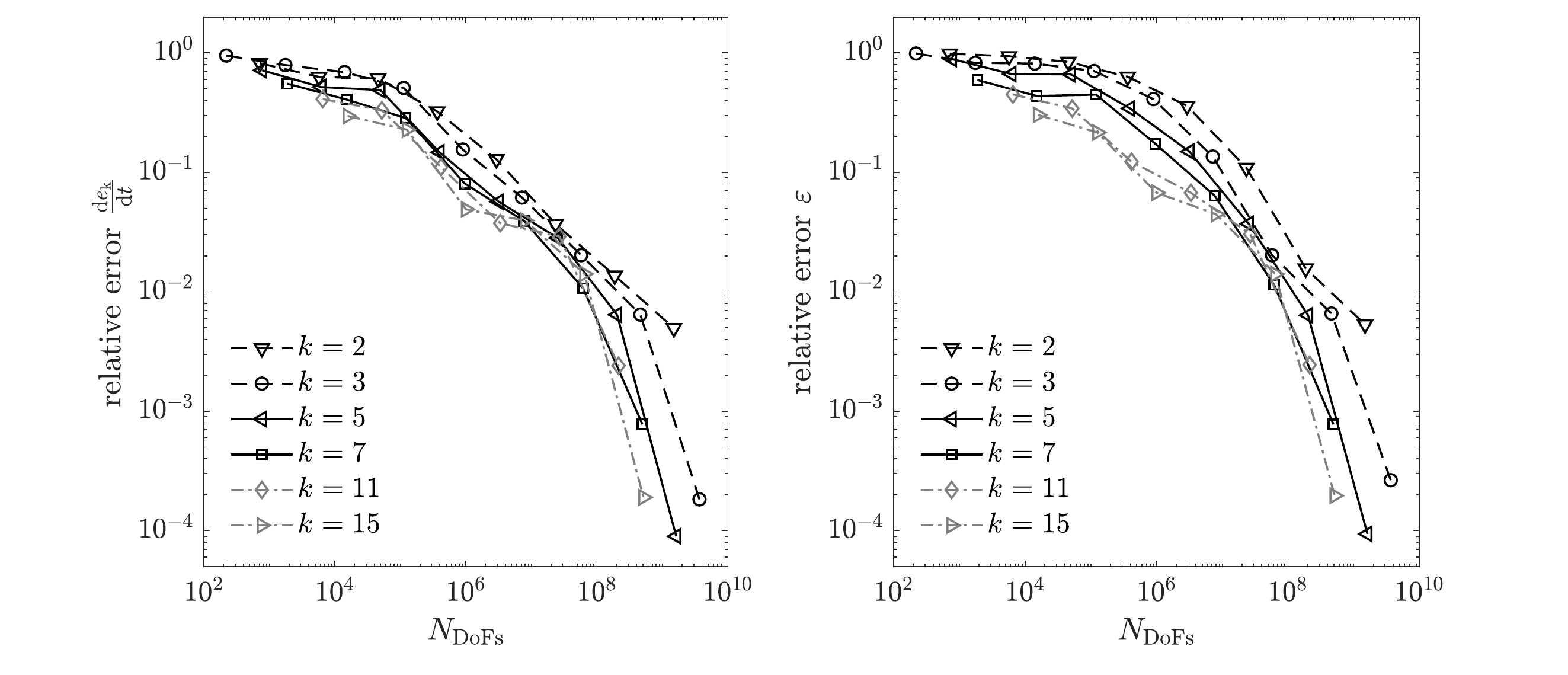}}
 \subfigure[Efficiency~$E_{h,k,\Delta t}$ of spatial and temporal discretization approach.]{
	 \includegraphics[width=1.0\textwidth]{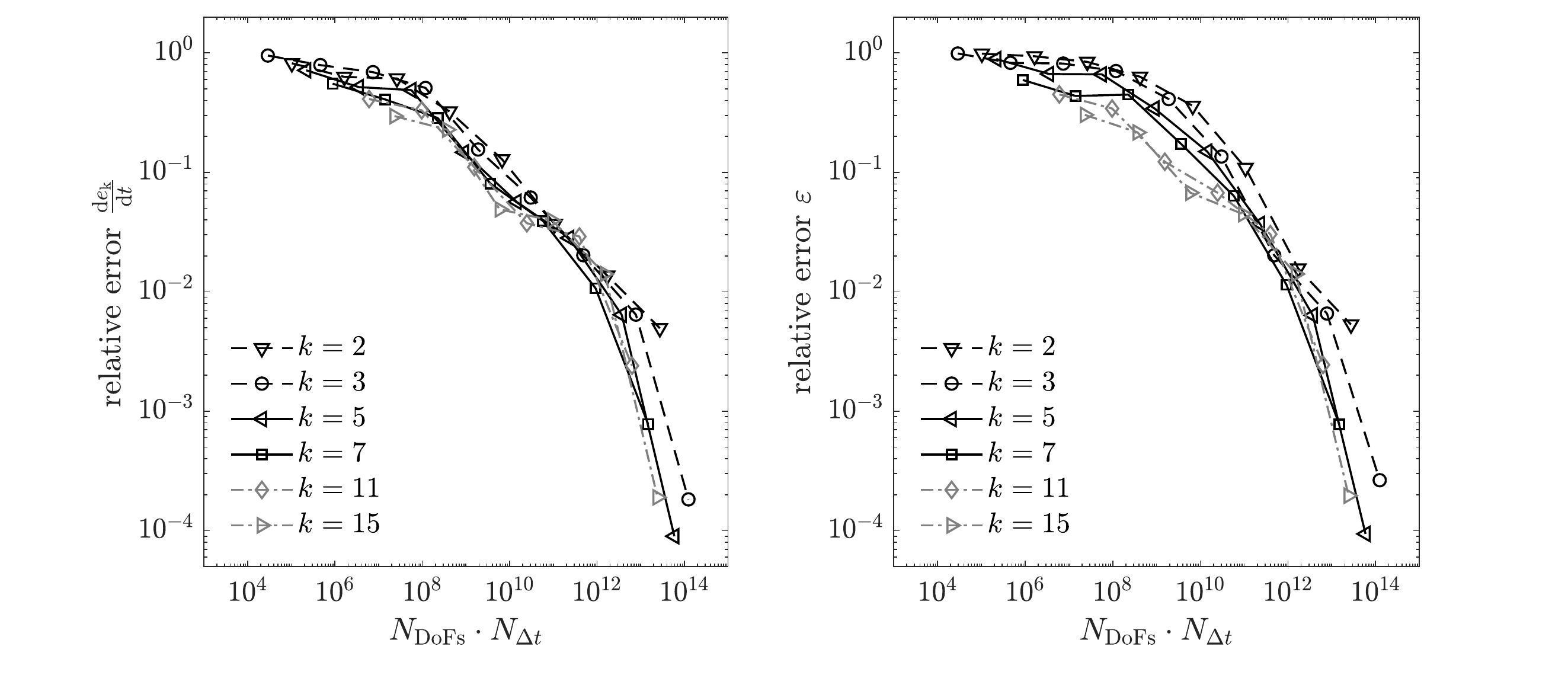}}
\caption{Efficiency of high-order spatial DG discretization in terms of accuracy versus number of unknowns as well as efficiency of temporal and spatial discretization approach in terms of accuracy versus number of unknowns times number of time steps.}
\label{fig:3D_Taylor_Green_Efficiency_Discretization}
\end{figure}

For a quantitive evaluation of the efficiency of high-order spatial discretization schemes, Figure~\ref{fig:3D_Taylor_Green_Efficiency_Discretization} shows the relative~$L^2$-errors of~$\frac{\mathrm{d}e_{\mathrm{k}}}{\mathrm{d}t}$ and~$\varepsilon$,
\begin{align*}
e_{e_{\mathrm{k}}}^2 = \frac{\int_{t=0}^{T} \left(  \frac{\mathrm{d}e_{\mathrm{k}}(t)}{\mathrm{d}t} - \frac{\mathrm{d}e_{\mathrm{k},\mathrm{ref}}(t)}{\mathrm{d}t} \right)^{2} \mathrm{d}t}{\int_{t=0}^{T} \left( \frac{\mathrm{d}e_{\mathrm{k},\mathrm{ref}}(t)}{\mathrm{d}t} \right)^{2} \mathrm{d}t} \;\; , \;\; 
e_{\varepsilon}^2 = \frac{\int_{t=0}^{T} \left(  \varepsilon(t) - \varepsilon_{\mathrm{ref}}(t) \right)^{2} \mathrm{d}t}{\int_{t=0}^{T} \left( \varepsilon_{\mathrm{ref}}(t) \right)^{2} \mathrm{d}t}   \;\; ,
\end{align*}
as a function of the number of unknowns~$N_{\mathrm{DoFs}}$ (efficiency of spatial discretization) and the number of unknowns times the number of time steps~$N_{\mathrm{DoFs}}\cdot N_{\Delta t}$ (efficiency of temporal and spatial discretization), where we use a double logarithmic scale for the plots as in Figure~\ref{fig:2D_Vortex_Problem}. The simulation with a resolution of~$1024^3$ and polynomial degree~$k=7$ is used for the calculation of errors. Regarding the spatial discretization, one can not expect optimal rates of convergence in the under-resolved regime. All polynomial degrees show a similar convergence trend where the rate of convergence is continuously increasing for increasing number of unknowns. With respect to the molecular dissipation~$\varepsilon$, the efficiency~$E_{h,k}$ of the spatial discretization approach improves systematically for increasing polynomial degree. Regarding the kinetic energy dissipation rate~$\frac{\mathrm{d}e_{\mathrm{k}}}{\mathrm{d}t}$, the differences in efficiency between high-order and low-order methods is reduced as compared to~$\varepsilon$, which is in agreement with the results shown in Figure~\ref{fig:3D_Taylor_Green_Kinetic_Energy_Dissipation}. For example, results for~$k=11,15$ are only slightly more accurate than for~$k=7$. Taking into account the time step size, the curves for high polynomial degrees are shifted towards the curve of the low polynomial degree~$k=2$ in agreement with the theoretical expectation according to the~$k^{1.5}$ relation, see equation~\eqref{CFL_Condition}. Regarding the kinetic energy dissipation rate~$\frac{\mathrm{d}e_{\mathrm{k}}}{\mathrm{d}t}$, one can not observe an advantage of very high-order methods~$k=11,15$ as compared to~$k=7$ in terms of the efficiency~$E_{h,k,\Delta t}$.

In order to verify that the temporal discretization errors are in fact negligible for Courant numbers close to the stability limit, we repeated the simulation for~$l=5$ and~$k=7$ with a resolution of $256^3$ (resulting in relative errors of approximately~$0.01$ for~$\frac{\mathrm{d}e_{\mathrm{k}}}{\mathrm{d}t}$ and~$\varepsilon$) for two smaller Courant numbers~$\mathrm{Cr}/2,\mathrm{Cr}/4$. Compared to the simulation with Courant number~$\mathrm{Cr}=0.125$, the relative change in the errors in not larger than~$1\% - 2 \%$ confirming the correctness of the chosen criterion for the selection of the time step size.

\subsection{Efficiency of solvers and preconditioners}\label{TGV_EfficiencySolvers}

\begin{table}[!h]
\caption{Iteration counts and relative share of computational costs per time step for pressure Poisson equation, projection equation, and Helmholtz equation of viscous step over a wide range of refinement levels and polynomial degrees. The Courant number is~$\mathrm{Cr} = 0.125$ for all simulations.}
\label{tab:IterationCountsAndRelativeCosts}
\renewcommand{\arraystretch}{1.1}
\begin{center}
\begin{small}
\subtable[Pressure step (Poisson equation)]{
\begin{tabular}{ccccccccccccccc}
\hline
\multicolumn{7}{c}{Iterations} & & \multicolumn{7}{c}{Relative costs}\\
\cline{1-7} \cline{9-15}
$l$ & \multicolumn{6}{c}{Polynomial degree~$k$} & & $l$ & \multicolumn{6}{c}{Polynomial degree~$k$}\\
\cline{2-7} \cline{10-15}
& 2 & 3 & 5 & 7 & 11 & 15 & & & 2 & 3 & 5 & 7 & 11 & 15 \\
\hline
0	 & $-$  & 2.0 & 2.0 & 2.0 & 2.0 & 2.0    & & 0	 & $-$  & 44\% & 32\% & 38\% & 53\% & 62\%\\
1	 & 0.0  & 6.2 & 6.3 & 7.0 & 8.7 & 10.8   & & 1	 & 15\% & 62\% & 54\% & 62\% & 77\% & 84\%\\
2	 & 5.0  & 6.3 & 6.1 & 7.0 & 8.4 & 11.0   & & 2	 & 35\% & 37\% & 33\% & 40\% & 47\% & 56\%\\
3	 & 8.9  & 7.0 & 6.3 & 7.0 & 8.6 & 11.5   & & 3	 & 55\% & 45\% & 38\% & 37\% & 43\% & 52\%\\
4	 & 9.6  & 7.1 & 6.3 & 7.0 & 8.9 & 11.7   & & 4	 & 58\% & 43\% & 33\% & 34\% & 43\% & 52\%\\
5	 & 9.3  & 7.1 & 6.3 & 6.9 & 8.7 & 11.3   & & 5	 & 36\% & 24\% & 27\% & 34\% & 42\% & 63\%\\
6	 & 9.4	& 7.1 & 6.8 & 6.9 & $-$ & $-$    & & 6	 & 26\% & 24\% & 30\% & 35\% & $-$  & $-$\\
7	 & 9.6 	& 7.0 & 6.5 & 6.5 & $-$ & $-$    & & 7	 & 27\% & 25\% & 30\% & 36\% & $-$  & $-$\\
8	 & 9.2 	& 7.2 & $-$ & $-$ & $-$ & $-$    & & 8	 & 28\% & 26\% & $-$  & $-$  & $-$  & $-$\\
\hline
\end{tabular}}

\subtable[Projection step (Projection equation including divergence and continuity penalty terms)]{
\begin{tabular}{ccccccccccccccc}
\hline
\multicolumn{7}{c}{Iterations} & & \multicolumn{7}{c}{Relative costs}\\
\cline{1-7} \cline{9-15}
$l$ & \multicolumn{6}{c}{Polynomial degree~$k$} & & $l$ & \multicolumn{6}{c}{Polynomial degree~$k$}\\
\cline{2-7} \cline{10-15}
& 2 & 3 & 5 & 7 & 11 & 15 & & & 2 & 3 & 5 & 7 & 11 & 15 \\
\hline
0	 & $-$  & 6.0  & 9.3  & 10.1 & 11.2 & 11.8  & & 0	 & $-$   & 29\% & 41\%  & 38\%  & 32\% & 26\%\\
1	 & 3.7  & 7.8  & 9.4  & 9.7  & 10.9 & 11.6  & & 1	 & 38\%  & 21\% & 27\%  & 23\%  & 16\% & 11\%\\
2	 & 7.7  & 8.3  & 9.6  & 10.2 & 11.3 & 12.5  & & 2	 & 31\%  & 32\% & 40\%  & 37\%  & 34\% & 30\%\\
3	 & 9.1  & 9.7  & 10.2 & 11.0 & 12.6 & 13.9  & & 3	 & 22\%  & 30\% & 34\%  & 38\%  & 38\% & 31\%\\
4	 & 9.9  & 10.3 & 10.9 & 11.6 & 12.9 & 14.0  & & 4	 & 23\%  & 31\% & 40\%  & 42\%  & 36\% & 30\%\\
5	 & 10.2 & 10.6 & 11.0 & 11.7 & 12.7 & 13.7  & & 5	 & 35\%  & 46\% & 47\%  & 41\%  & 36\% & 22\%\\
6	 & 10.4	& 10.5 & 10.9 & 11.3 & $-$  & $-$   & & 6	 & 45\%  & 47\% & 43\%  & 40\%  & $-$  & $-$\\
7	 & 10.1	& 10.3 & 10.4 & 10.4 & $-$  & $-$   & & 7	 & 42\%  & 45\% & 42\%  & 39\%  & $-$  & $-$\\
8	 & 9.4 	& 9.6  & $-$  & $-$  & $-$  & $-$   & & 8	 & 39\%  & 42\% & $-$   & $-$   & $-$  & $-$\\
\hline
\end{tabular}}

\subtable[Viscous step (Helmholtz equation)]{
\begin{tabular}{ccccccccccccccc}
\hline
\multicolumn{7}{c}{Iterations} & & \multicolumn{7}{c}{Relative costs}\\
\cline{1-7} \cline{9-15}
$l$ & \multicolumn{6}{c}{Polynomial degree~$k$} & & $l$ & \multicolumn{6}{c}{Polynomial degree~$k$}\\
\cline{2-7} \cline{10-15}
& 2 & 3 & 5 & 7 & 11 & 15 & & & 2 & 3 & 5 & 7 & 11 & 15 \\
\hline
0	 & $-$  & 2.0 & 2.5 & 3.0 & 2.9 & 2.9   & & 0	 & $-$  & 21\% & 22\% & 20\% & 12\% & 9\%\\
1	 & 2.0  & 2.2 & 3.0 & 3.0 & 3.0 & 2.9	& & 1	 & 37\% & 14\% & 15\% & 12\% & 6\%  & 4\%\\
2	 & 3.0  & 3.0 & 3.0 & 3.1 & 3.7 & 3.8   & & 2	 & 29\% & 26\% & 22\% & 18\% & 15\% & 12\%\\
3	 & 3.0  & 3.0 & 3.7 & 3.8 & 3.8 & 4.7   & & 3	 & 16\% & 19\% & 23\% & 22\% & 16\% & 14\%\\
4	 & 3.0  & 3.8 & 3.8 & 3.8 & 4.9 & 5.8   & & 4	 & 15\% & 23\% & 23\% & 21\% & 18\% & 15\%\\
5	 & 3.8  & 3.8 & 3.9 & 4.4 & 5.3 & 5.9   & & 5	 & 25\% & 26\% & 23\% & 21\% & 19\% & 12\%\\
6	 & 3.9	& 4.1 & 4.3 & 4.7 & $-$ & $-$   & & 6	 & 25\% & 25\% & 23\% & 22\% & $-$  & $-$\\
7	 & 4.6  & 4.4 & 4.5 & 4.5 & $-$ & $-$   & & 7	 & 27\% & 27\% & 24\% & 22\% & $-$  & $-$\\
8	 & 5.1 	& 4.6 & $-$ & $-$ & $-$ & $-$   & & 8	 & 30\% & 28\% & $-$  & $-$  & $-$  & $-$\\
\hline
\end{tabular}}

\end{small}
\end{center}
\renewcommand{\arraystretch}{1}
\end{table}

We analyze the efficiency of the solution of linear systems of equations in terms of the number of iterations required to solve the pressure Poisson equation~\eqref{DualSplitting_Pressure_MatrixForm}, the projection equation~\eqref{DualSplitting_Projection_MatrixForm}, and the Helmholtz equation~\eqref{DualSplitting_ViscousStep_MatrixForm}. Results are presented in Table~\ref{tab:IterationCountsAndRelativeCosts}. For the pressure Poisson equation, mesh-independent convergence is achieved for all polynomial degrees, i.e., the number of iterations approaches a constant value for large spatial refinement levels~$l$, when using the geometric multigrid method with polynomial Chebyshev smoother. The lowest number of iterations is obtained for moderate polynomial degrees~$k=5$, while the number of iterations slightly increases for low polynomial degrees and very high polynomial degrees. Mesh-independent convergence is also obtained for the projection equation and the Helmholtz equation of the viscous step where the inverse mass matrix operation is used as a preconditioner. These results demonstrate that the inverse mass matrix is a very effective preconditioner for these equations in case of small time step sizes. In the next section we will show that the inverse mass matrix operation is also very efficient in terms of computational costs (as expensive as scaling a vector by a diagonal matrix), making this preconditioner also the most efficient preconditioner for the Helmholtz equation and the projection equation. As explained in Section~\ref{Efficiency}, it is not only the number of global iterations determining the efficiency of a preconditioner but rather the equivalent computational costs for applying this preconditioner. In this respect, the geometric multigrid preconditioner applied to the solution of the pressure Poisson equation is significantly more complex than the inverse mass matrix preconditioner used for the other equations. Hence, we also include the relative share of the overall computational costs per time step for the three main substeps of the splitting scheme. The numbers in Table~\ref{tab:IterationCountsAndRelativeCosts} highlight that the three substeps are well-balanced and that there is no obvious bottleneck in the solver. The Helmholtz solver is cheapest in comparison due to the small number of iterations. Depending on the spatial resolution parameters~$l$ and~$k$, it is either the pressure Poisson equation or the projection equation that forms the most expensive substep. The geometric multigrid preconditioner with Chebyshev smoother is more expensive than the inverse mass matrix operator in terms of an effective number of matrix--vector products, which is offset by the lower polynomial degree for the pressure, $k_p=k_u-1$, and the fact that pressure is a scalar quantity. In terms of wall time, the relative share of the convective step is between~$1\%$ and~$7\%$ of the overall computational costs per time step, most often around~$3\%-4\%$, despite the fact that an increased number of quadrature points is used according to the~$3/2$-rule. Hence, the convective step is negligible in terms of performance. In the next section, we investigate the computational efficiency of the performance relevant operations applied within the linear solvers.

\subsection{Efficiency of matrix-free implementation and potential for optimizations}\label{TGV_EfficiencyImplementation}
\begin{table}
\renewcommand{\arraystretch}{1.1}
\begin{center}
\begin{small}
\begin{tabular}{llll}
\hline
\multicolumn{2}{l}{Processor} & \multicolumn{2}{l}{Memory and Caches}\\
\hline
Processor type &  Intel Xeon E5-2697 v3 & Shared memory per node & 64 GByte\\
Frequency & 2.6 GHz & Bandwidth to memory per node & 137 GByte/sec\\
Cores per node & 28 & Cache sizes for level 1 $\vert$ 2 $\vert$ 3 & 32 $\vert$ 256 $\vert$ $2\cdot 35000$  kByte \\
\hline
\end{tabular}
\end{small}
\end{center}
\renewcommand{\arraystretch}{1}
\caption{Performance characteristics of SuperMUC Phase 2. L1 and L2 caches are specified per core, the shared L3 cache for the whole node.}
\label{SuperMUC_Phase2}
\end{table}

The pressure Poisson operator, the Helmholtz operator, and the projection operator are the performance relevant operators for the Navier--Stokes solver considered in this work. Again, we emphasize that the evaluation of the nonlinear convective term and the increased computational costs per degree of freedom due to an increased number of quadrature points is less of a concern since this operator is applied only once within each time step while the above operators are applied multiple times during the solution of linear systems of equations. Preconditioning of the pressure Poisson equation is based on geometric multigrid methods with a polynomial Chebyshev smoother which involves repeated applications of the discretized Laplace operator, see~\cite{Adams03}. The Helmholtz equation and the projection equation are preconditioned by the inverse mass matrix operator. In the following, we analyze the performance of the inverse mass matrix operation in addition to the three operators constituting the linear systems of equations~\eqref{DualSplitting_Pressure_MatrixForm},~\eqref{DualSplitting_Projection_MatrixForm}, and~\eqref{DualSplitting_ViscousStep_MatrixForm}.

\begin{figure}
 \centering 
\includegraphics[width=1.0\textwidth]{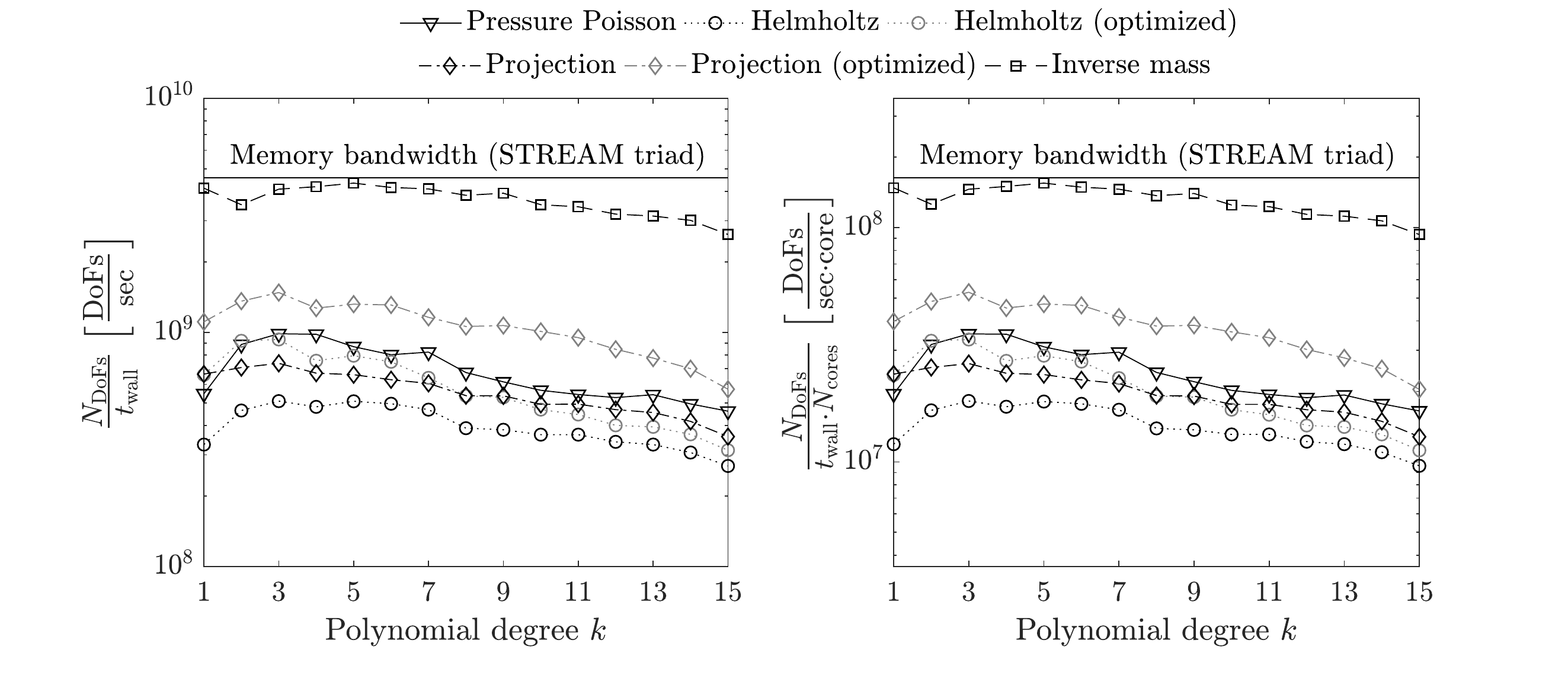}
\caption{Performance results of matrix-free operator evaluation for pressure Poisson operator, Helmholtz operator, projection operator, and inverse velocity mass matrix operator.}
\label{fig:Efficiency_matrix_free}
\end{figure}

Performance measurements are carried out on SuperMUC Phase 2 in Garching, Germany based on an Intel Haswell processor architecture with AVX2 instruction set extension. The main performance characteristics of this system are summarized in Table~\ref{SuperMUC_Phase2}. We measure the performance of the matrix-free operator evaluation for polynomial degrees in the range~$1\leq k \leq 15$ in terms of throughput~$N_{\mathrm{dofs}}/t_{\mathrm{wall}}$ in~$\mathrm{DoFs}/\mathrm{sec}$ as well as the efficiency of the implementation given by~$E_{\bm{A}\bm{x}}=N_{\mathrm{dofs}}/(t_{\mathrm{wall}}\cdot N_{\mathrm{cores}})$ in~$\mathrm{DoFs}/(\mathrm{sec}\cdot\mathrm{core})$, where~$N_{\mathrm{cores}}$ denotes the number of cores (MPI processes). The simulations are run on a fully loaded node with 28 MPI processes. The problem size is chosen large enough to avoid cache effects, i.e., we use refinement level~$l=7$ for~$k=1$,~$l=6$ for~$2\leq k \leq 3$, refinement level~$l=5$ for~$4\leq k \leq 7$, and refinement level~$l=4$ for~$8\leq k \leq 15$. The wall time~$t_{\mathrm{wall}}$ is measured as the wall time averaged over 100 operator evaluations taking the minimum out of 10 consecutive runs. Results of this performance experiment are presented in Figure~\ref{fig:Efficiency_matrix_free}.
For the inverse mass matrix operator, the throughput is close to the memory bandwidth limit. The theoretical limit is~$137 \;\mathrm{GByte}/\mathrm{sec}$ as listed in Table~\ref{SuperMUC_Phase2}, but the achievable bandwidth for the STREAM triad benchmark is only~$\approx 110 \;\mathrm{GByte}/\mathrm{sec}$, corresponding to a throughput of~$110\cdot 10^9/(8\cdot 3)\; \mathrm{DoFs}/\mathrm{sec}= 4.6\cdot 10^9\;\mathrm{DoFs}/\mathrm{sec}$, where the factors~$8$ and~$3$ account for the fact that we consider double precision with~$8$ Bytes per DoF and that we have to read two vectors from memory (input and output, the latter due to the read-for-ownership policy~\cite{Hager2010}) and write one vector to memory. 
This result demonstrates that the inverse mass matrix operator is memory bandwidth bound and is as expensive as applying a diagonal mass matrix although the consistent mass matrix is block diagonal and is inverted exactly in our implementation. The pressure Poisson operator, the Helmholtz operator and the projection operator achieve a throughput of~$3\cdot 10^8 - 1\cdot 10^9 \;\mathrm{DoFs}/\mathrm{sec}$ on all 28 cores or~$1\cdot 10^7 - 4\cdot 10^7 \;\mathrm{DoFs}/(\mathrm{sec}\cdot\mathrm{core})$. The best performance is achieved for polynomial degrees in the range~$2\leq k \leq 7$ where the throughput is almost independent of~$k$. For higher polynomial degrees, the efficiency slightly decreases related to the increased complexity of cell integrals as compared to face integrals where the former dominate the overall costs for large~$k$. Besides the increased computational work, higher polynomial degrees beyond~$k=10$ also involve larger temporary arrays for the tensor product kernels that cannot solely be handled by the fast L1 and L2 caches, slightly reducing throughput~\cite{Kronbichler2017b}. Moreover, for the same number of unknowns and compute cores, MPI communication overhead increases for large polynomial degrees due to the smaller number of elements per core as compared to low-order methods and the increased surface-to-volume ratio relevant for MPI ghost layer communication~\cite{Hager2010}. For a high workload (number of elements) per core this effect is negligible but it becomes relevant if the aim is to minimize the wall time as in strong scaling experiments where the number of elements per core becomes small. 
\begin{table}
\renewcommand{\arraystretch}{1.1}
\begin{center}
\begin{small}

\subtable[Standard implementation]{
\begin{tabular}{cccccccc}
\hline
& & &\multicolumn{2}{c}{Projection step}  & & \multicolumn{2}{c}{Viscous step} \\
\cline{4-5} \cline{7-8}
$k$ & $l$ & computational costs & solver & mat-vec & & solver & mat-vec\\
\hline
7 & 4 & $3.3 \cdot 10^{1}$ CPUh & 42\% & 24\% & & 21\% & 13\% \\
  & 5 & $5.8 \cdot 10^{2}$ CPUh & 41\% & 23\% & & 21\% & 13\%\\
\hline
\end{tabular}
}

\subtable[Optimized implementation of projection operator and Helmholtz operator]{
\begin{tabular}{cccccccc}
\hline
& & &\multicolumn{2}{c}{Projection step}  & & \multicolumn{2}{c}{Viscous step} \\
\cline{4-5} \cline{7-8}
$k$ & $l$ & computational costs & solver & mat-vec & & solver & mat-vec\\
\hline
7 & 4 & $2.9 \cdot 10^{1}$ CPUh & 38\% & 17\% & & 20\% & 12\% \\
  & 5 & $5.0 \cdot 10^{2}$ CPUh & 36\% & 15\% & & 20\% & 11\% \\
\hline
\end{tabular}
}
\end{small}
\end{center}
\renewcommand{\arraystretch}{1}
\caption{Comparison of standard and optimized implementations: Overall computational costs in CPUh (without setup costs) as well as relative share of computational costs per time step for projection step and viscous step (further separated into costs for the whole conjugate gradient solver and for the matrix-free operator evaluation, labeled mat-vec, only).}
\label{tab:Standard_vs_optimized_projection_operator}
\end{table}

Differences in throughput between the pressure Poisson, Helmholtz, and projection operators are within a factor of 2 and are related to different complexities of these operators, e.g., the Helmholtz operator and the projection operator are applied to a vectorial quantity as opposed to the scalar pressure Poisson operator and are composed of several basic operators. Due to the highly optimized implementation of the pressure Poisson operator, we focus on possible performance improvements for the projection and Helmholtz operators. In Table~\ref{tab:Standard_vs_optimized_projection_operator}, the relative share of the overall computational costs for applying the projection operator and the Helmholtz operator are listed for two simulations with degree~$k=7$ and refinement levels~$l=4,5$. These results reveal that a significant part of the computational costs of the projection and the Helmholtz solvers is actually not spent in the matrix-vector products but in the three vector updates, the two inner products, the residual norm, and the inverse mass matrix preconditioner within the conjugate gradient algorithm. Accordingly, due to the high optimization level of our matrix-free implementation, further performance improvements of the operators already have a limited impact on the overall performance.

For the projection operator and the Helmholtz operator, there is some potential for further optimization. In the standard implementation of the projection operator, the mass matrix operator, the divergence penalty operator, and the continuity penalty operator are applied operator-by-operator for reasons of code modularity and flexibility. Combining these three operators to one single operator on the one hand and improving the MPI communication for the face integrals of the continuity penalty operator~\cite{Kronbichler2017b} on the other hand (both measures yield a similar improvement in performance), the performance of the projection operator can be further improved by a factor of~$1.6-2.1$ as shown in Figure~\ref{fig:Efficiency_matrix_free}. For the Helmholtz operator, the performance can be improved by a factor of~$1.2-2.0$ by combining the mass matrix and viscous operators as well as avoiding unnecessary if-else branches and function calls inside the inner loop over the quadrature points. Using the optimized implementation, the overall costs can be reduced by approximately~$11\%-14\%$ as shown in Table~\ref{tab:Standard_vs_optimized_projection_operator}. While the relative share of the matrix-free operator evaluation within the CG algorithm for the projection equation is approximately~$55\%$ for the standard implementation, it is only approximately~$40\%-45\%$ for the optimized projection operator. Similarly, the relative costs for the Helmholtz operator within the CG algorithm are approximately~$ 65\%$ for the standard implementation and only~$55\%-60\%$ for the optimized implementation. Hence, slight improvements in the performance of the projection operator or the Helmholtz operator will only have a small impact on the overall costs. For the performance numbers and overall computational costs presented in the following, the standard implementation has been used since this is the performance that can be expected for a straightforward, modular implementation using the generic matrix-free implementation.

\subsection{Evaluation of overall efficiency}\label{TGV_OverallEfficiency}
We now face the question raised in the beginning: Does the efficiency of the numerical method defined as the ratio of accuracy and computational costs increase for increasing polynomial degree~$k$?

In Table~\ref{tab:WallTimesAndCosts}, the wall time as well as the overall computational costs are listed for several refinement levels~$l$ and for polynomial degrees~$k=3,5,7$. As usual, the overall computational costs are the costs for the whole time loop (including postprocessing after each time step) but without setup costs. The number of compute nodes (cores) is chosen such that the wall time does not exceed the wall time limit of 48 hours. The overall computational costs increase by a factor close to~$16$ from one refinement level to the next demonstrating the close to optimal complexity as well as excellent parallel scalability of our algorithm. Before considering the overall efficiency by taking into account the error of the individual simulations, we compare our solver to performance results published in the literature for high-order compressible DG schemes. In~\cite{Wiart14}, an effective spatial resolution of~$256^3$ realized by~$l=6$ and~$k=3$ has been calculated, with overall computational costs of~$6.4$ kCPUh. Since the final time is~$T=10$ in~\cite{Wiart14}, the computational costs would be approximately~$2.3 \cdot 10^{2}\;\mathrm{CPUh}/2=1.15 \cdot 10^{2}$ CPUh for the same time interval and the present solution approach (see Table~\ref{tab:WallTimesAndCosts}), yielding a performance improvement of our approach by factor of~$6.4 \cdot 10^{3}/1.15 \cdot 10^{2} \approx 56$. A very high-order method with~$k=15$ and an effective resolution of~$64^3$ has been simulated in~\cite{Gassner2013}, where wall times in the range~$10 \;\mathrm{min}-1.3\;\mathrm{h}$ are specified (on 64 processors), resulting in overall computational costs of~$10.7-83.2$ CPUh. The lower value is obtained for a filtering approach and the upper value for a more accurate overintegration strategy that is also used in the present work for the convective term. Again, the end time is~$T=10$ in~\cite{Gassner2013}. For the present solver, the computational costs are~$3.8\;\mathrm{CPUh}/2=1.9$ CPUh (computed on a single core due to the small number of elements) yielding a speed-up by a factor of~$6-44$. As shown in~\cite{Fehn18}, the present discretization approach is competitive to the compressible DG code used in~\cite{Gassner2013} in terms of accuracy. Taking into account that the results have been obtained on different hardware, we expect that the present incompressible DG solver allows an improvement in overall performance by at least one order of magnitude as compared to state-of-the-art high-order compressible DG solvers. We also note that even larger time step sizes could be used for the present approach according to Table~\ref{CriticalCourantNumbers} further reducing computational costs. Moreover, implementation techniques such as the operator-integration-factor splitting~\cite{Maday1990} could likely reduce costs by another factor of 2.

Our approach is also highly competitive to other discretization schemes such as finite volume methods or continuous finite element methods. For example, a wall time of~$63$ min on an Intel SandyBridge system with 16 cores are specified in~\cite{Schranner2016} for a finite volume discretization with a resolution of~$64^3$ and an end time of~$T=10$ resulting in computational costs of approximately~$1.7\cdot 10^1$ CPUh. For the~$\mathrm{AVM}^4$ turbulence model~\cite{Rasthofer2013} based on a low-order continuous finite element discretization, we measured computational costs of~$2.9 \cdot 10^{1}$ CPUh on an Intel Haswell system for a resolution of~$64^3$ and end time~$T=20$. Selecting ~$l=4$ and~$k=3$ (resolution~$64^3$) for the present discretization approach with~$T=20$ results in costs of~$7.5 \cdot 10^{-1}$ CPUh (see Table~\ref{tab:WallTimesAndCosts}), so that our solver achieves an improvement in computational costs as compared to the finite volume approach~\cite{Schranner2016} and the finite element approach~\cite{Rasthofer2013} by a factor of 45 and 39, respectively. These numbers show that the use of efficient high-order DG methods allow to signficantly improve the performance as compared to state-of-the-art numerical methods for turbulent flows that are not specifically optimized for performance.

\begin{table}
\renewcommand{\arraystretch}{1.1}
\begin{center}
\begin{tabular}{llllllll}
\hline
$k$ & $l$ & resolution & $N_{\mathrm{DoFs}}$ & $N_{\Delta t}$ & $t_{\mathrm{wall}}$ [s] & $N_{\mathrm{cores}}$ & $t_{\mathrm{wall}}\cdot N_{\mathrm{cores}}$ [CPUh]\\
\hline
3 & 4 & $64^3$   & $9.0\cdot 10^{5}$ & 2118  & $9.6\cdot 10^{1}$ & $28$                 & $7.5 \cdot 10^{-1}$ \\
  & 5 & $128^3$  & $7.2\cdot 10^{6}$ & 4235  & $1.7\cdot 10^{3}$ & $28$                 & $1.3 \cdot 10^{1}$  \\
  & 6 & $256^3$  & $5.7\cdot 10^{7}$ & 8469  & $1.5\cdot 10^{4}$ & $28\cdot 2 = 56$     & $2.3 \cdot 10^{2}$  \\
  & 7 & $512^3$  & $4.6\cdot 10^{8}$ & 16937 & $3.1\cdot 10^{4}$ & $28\cdot 16 = 448$   & $3.8 \cdot 10^{3}$  \\
  & 8 & $1024^3$ & $3.7\cdot 10^{9}$ & 33874 & $6.4\cdot 10^{4}$ & $28\cdot 128 = 3584$ & $6.4 \cdot 10^{4}$  \\
\hline
5 & 4 & $96^3$  & $3.2\cdot 10^{6}$ & 4556  & $7.9\cdot 10^{2}$ & $28$ 			      & $6.2 \cdot 10^{0}$ \\
  & 5 & $192^3$ & $2.5\cdot 10^{7}$ & 9111  & $1.4\cdot 10^{4}$ & $28$ 			      & $1.1 \cdot 10^{2}$ \\
  & 6 & $384^3$ & $2.0\cdot 10^{8}$ & 18222 & $3.2\cdot 10^{4}$ & $28\cdot 8 = 224$   & $2.0 \cdot 10^{3}$ \\
  & 7 & $768^3$ & $1.6\cdot 10^{9}$ & 36443 & $6.4\cdot 10^{4}$ & $28\cdot 64 = 1792$ & $3.2 \cdot 10^{4}$ \\
\hline
7 & 4 & $128^3$  & $7.7\cdot 10^{6}$ & 7546  & $4.3\cdot 10^{3}$ & $28$ 			     & $3.3 \cdot 10^{1}$ \\
  & 5 & $256^3$  & $6.2\cdot 10^{7}$ & 15092 & $1.9\cdot 10^{4}$ & $28\cdot 4 = 112$     & $5.8 \cdot 10^{2}$ \\
  & 6 & $512^3$  & $4.9\cdot 10^{8}$ & 30184 & $7.6\cdot 10^{4}$ & $28\cdot 16 = 448$    & $9.4 \cdot 10^{3}$ \\
  & 7 & $1024^3$ & $3.9\cdot 10^{9}$ & 60367 & $1.0\cdot 10^{5}$ & $28\cdot 192 = 5376 $ & $1.6 \cdot 10^{5}$ \\
\hline
\end{tabular}
\end{center}
\renewcommand{\arraystretch}{1}
\caption{Performance results for polynomial degrees~$k=3,5,7$ and refinement levels~$l=4,...,8$. The time interval is~$0\leq t \leq T$ with end time~$T=20 T_0=20\frac{L}{U_0}$. The Courant number is~$\mathrm{Cr}=0.125$ for all computations.}
\label{tab:WallTimesAndCosts}
\end{table}

\begin{figure}[!ht]
 \centering 
 \subfigure[Comparison of overall efficiency for polynomial degrees~$k=2,3,5,7,11,15$.]{
	 \includegraphics[width=0.9\textwidth]{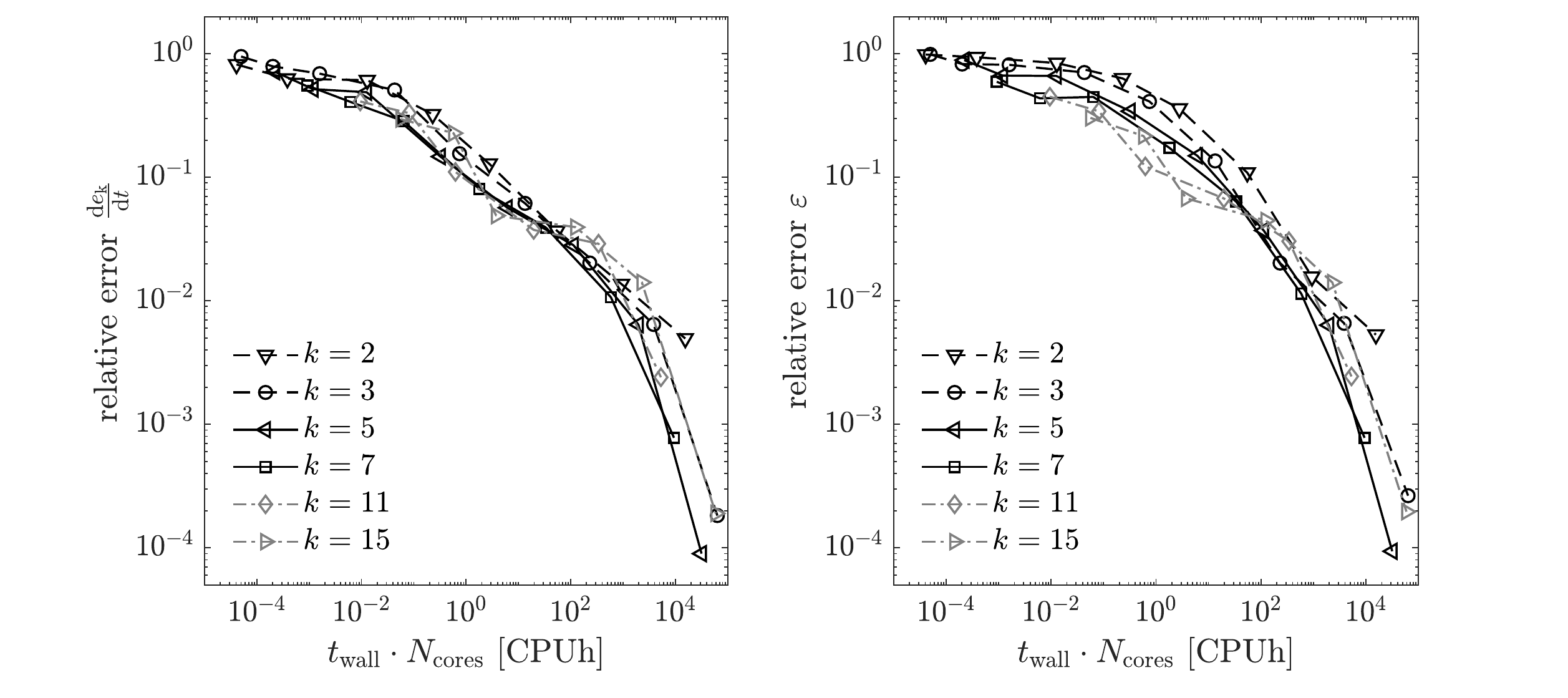}\label{subfig:all_k}}
 \subfigure[Efficiency of most efficient method with~$k=7$ compared to lower order method with~$k=2$.]{
	 \includegraphics[width=0.9\textwidth]{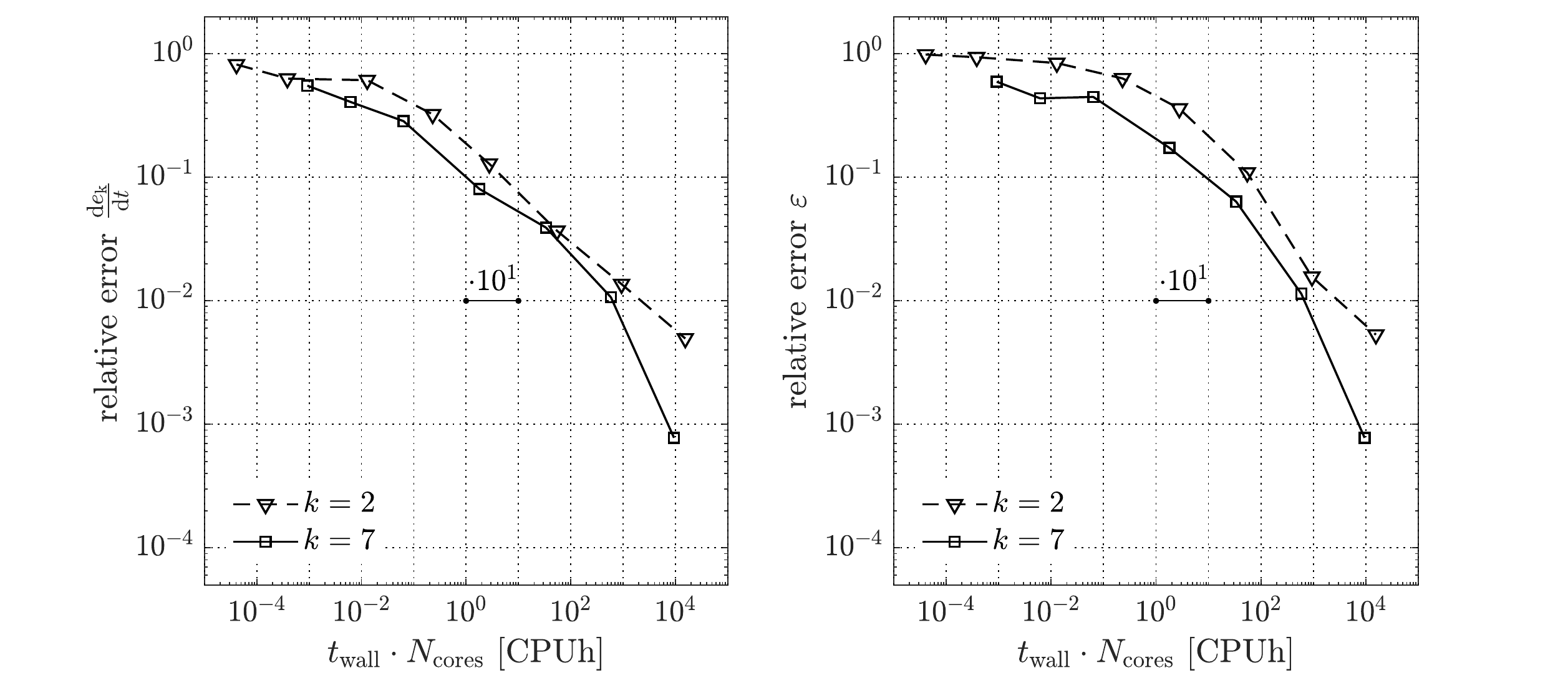}\label{subfig:k2_k7_only}}
\caption{Overall efficiency~$E$ of computational approach in terms of error versus computational costs.}
\label{fig:3D_Taylor_Green_Overall_Efficiency}
\end{figure}

Finally, we consider the overall efficiency~$E$ in Figure~\ref{fig:3D_Taylor_Green_Overall_Efficiency}~\subref{subfig:all_k}, where the error is plotted over the computational costs for polynomial degrees~$k=2,3,5,7,11,15$. The best efficiency is obtained for moderate polynomial degrees~$k=5,7$, for which the efficiency is significantly improved as compared to the lower polynomial degrees~$k=2,3$. Although the curves for different polynomial degrees appear to be very close to each other, we note that the improvement in overall efficiency is up to one order of magnitude for~$k=7$ as compared to~$k=2$. To highlight this gain in performance for high-order methods, the overall efficiency for polynomial degree~$k=7$ is explicitly compared to~$k=2$ in Figure~\ref{fig:3D_Taylor_Green_Overall_Efficiency}~\subref{subfig:k2_k7_only}. While there is an advantage of very high polynomial degrees~$k=11,15$ regarding the dissipation rate~$\varepsilon$ for some refinement levels, these polynomial degrees are less efficient than~$k=5,7$ with respect to the kinetic energy dissipation rate~$\frac{\mathrm{d}e_{\mathrm{k}}}{\mathrm{d}t}$ which might be considered as the more relevant quantity from an application point of view. The reasons for the observed behavior have already been indicated in Sections~\ref{TGV_EfficiencySolvers} and~\ref{TGV_EfficiencyImplementation}: Although our approach achieves close to optimal performance with respect to preconditioning aspects (efficiency~$E_{\bm{x}=\bm{A}^{-1}\bm{b}}$) and implementation issues (efficiency~$E_{\bm{A}\bm{x}}$), a slight increase in iterations counts as well as a slightly reduced throughput of the matrix-free evaluation for very high polynomial degrees in addition to time step limitations related to the CFL condition renders very high-order methods less efficient than moderate high-order methods for problems that lack optimal rates of convergence in space such as under-resolved, turbulent flows. Similar results for the Taylor--Green vortex problem considering the error versus the computational costs are shown in~\cite{Wang2013} where the TauBench code is used in order to normalize the computational costs to obtain work units. Normalizing the computational costs as suggested in~\cite{Wang2013} reveals that the present approach allows to reduce the computational costs by a factor of~$10-100$ for a given level of accuracy against the methods analyzed in~\cite[Figure~25]{Wang2013}.

\section{Summary and outlook}\label{Summary}
We have presented performance results for a high-order DG, incompressible Navier--Stokes solver for under-resolved turbulent flows by the example of the 3D Taylor--Green vortex problem. For preconditioning of symmetric, positive definite systems of equations we use the geometric multigrid method with polynomial Chebyshev smoothing and the inverse mass matrix for the projection step and the viscous step resulting in mesh-independent convergence rates. The implementation is based on a high-performance matrix-free implementation and all components of the solver and preconditioner are implemented in a matrix-free way. Efficiency is defined as the ratio of accuracy and computational costs. A methodology has been introduced to systematically analyze the efficiency of such an approach with respect to high polynomial degrees, identifying the efficiency of the spatial and temporal discretization, the efficiency of the solver, and the efficiency of the implementation as the main factors. A detailed performance analysis of all relevant components has been carried out over a wide range of mesh refinement levels and polynomial degrees. To the best of the authors' knowledge, such an analysis has not been published to-date for high-order DG discretizations of the incompressible Navier--Stokes equations applied to turbulent flows.

The main results can be summarized as follows: The lack of optimal rates of convergence in the under-resolved regime raises the question regarding the efficiency of spectral element type methods or, more concretely, the question regarding the optimal polynomial degree providing the best efficiency. In fact, an advantage of high-order methods in terms of accuracy can be observed for the same number of unknowns. However, attention should be paid to the more relevant metric of accuracy versus computational costs. Despite the fact that our approach achieves optimal complexity and that our implementation is highly optimized for modern computer architectures, demonstrating improved efficiency of high-order methods for problems as simple as the 3D Taylor--Green vortex problem is a challenging task. The most efficient results are obtained for moderate polynomial degrees~$k=5,7$. For higher polynomial degrees~$k=11,15$ the overall efficiency could not be further improved. The three main components explaining this behavior are: (i) restrictions of the time step size according to the CFL condition in case of an explicit treatment of the convective term reducing the efficiency of high-order methods, (ii) a slight increase in iteration counts for solving the linear systems for very high-order methods, (iii) and a slight reduction in the computational efficiency of the matrix-free evaluation of discretized operators for very high-order methods due to an increasing arithmetic intensity.

Although we focused on the geometrically simple Taylor--Green vortex problem, our approach is generic and can be applied to arbitrary geometries. The efficiency of the matrix-free implementation directly translates to more complex, non Cartesian meshes. The accuracy and efficiency of the spatial discretization approach when applied to wall-bounded turbulent flows and problems involving singularities is an open issue to be analyzed. According to the present study, a significant advantage of high-order methods in terms of accuracy is a necessary prerequisite for high-order methods to be more efficient even for close to optimal implementations. The efficiency might be further improved by considering implicit time integration for the convective term to avoid the CFL restriction. The development of efficient, matrix-free preconditioners in the convection-dominated regime poses a major challenge and is considered as part of future work.

Compared to results published in the literature for state-of-the-art, compressible, high-order DG solvers, our solution approach allows to reduce the wall time by more than one order of magnitude for the same spatial resolution. The main differences between compressible and incompressible DG solvers are the following: Explicit time stepping methods often used for compressible Navier--Stokes solvers might result in very restrictive time step sizes for fine spatial resolutions. The spatial derivative operators are more complex for the compressible Navier--Stokes equation, while one only has to solve a set of symmetric, positive definite linear systems of equations in case of the present incompressible Navier--Stokes solver. The required number of quadrature points is larger for the compressible than for the incompressible Navier--Stokes equations in order to avoid aliasing effects. Moreover, the increased number of quadrature points directly influences the overall efficiency in case of compressible solvers, while the exact quadrature used for the nonlinear convective term does not deteriorate the performance of the present solver. Finally, the implementation might have a significant impact on the observed differences in performance. As part of future work, we plan to implement a compressible high-order DG code within our high-performance, matrix-free implementation framework in order to clarify this issue.

\appendix

\section*{Acknowledgments}
The research presented in this paper was partly funded by the German Research Foundation (DFG) under the project ``High-order discontinuous Galerkin for the EXA-scale'' (ExaDG) within the priority program ``Software for Exascale Computing'' (SPPEXA), grant agreement no. KR4661/2-1 and WA1521/18-1. The authors gratefully acknowledge the Gauss Centre for Supercomputing e.V.~(\texttt{www.gauss-centre.eu}) for funding this project by providing computing time on the GCS Supercomputer SuperMUC at Leibniz
Supercomputing Centre (LRZ, \texttt{www.lrz.de}) through project id pr83te.

\bibliography{paper}

\end{document}